\documentclass[trackchanges, twocolumn]{aastex701}

\usepackage[T1]{fontenc}
\usepackage{amsmath}
\colorlet{RED}{red}

\begin{document}

\title{An End-to-End Numerical Framework for Tidal Disruption Events with \texttt{AthenaK}}

\author[orcid=0000-0003-0292-2773]{Hong-Xuan Jiang}
\email{hongxuan\_jiang@sjtu.edu.cn}
\affiliation{Tsung-Dao Lee Institute, Shanghai Jiao Tong University, Shengrong Road 520, Shanghai, 201210, China}
\affiliation{The Hong Kong Institute for Astronomy and Astrophysics, The University of Hong Kong, Pokfulam Road, Hong Kong, China}

\author[orcid=0009-0007-2022-721X]{Mengqi Yang}
\email{pachelbel@sjtu.edu.cn}
\affiliation{Tsung-Dao Lee Institute, Shanghai Jiao Tong University, Shengrong Road 520, Shanghai, 201210, China}
\affiliation{Key Laboratory for Laser Plasmas (MoE) and School of Physics and Astronomy, Shanghai Jiao Tong University, Shanghai 200240, People’s Republic of China}
\affiliation{Collaborative Innovation Center of IFSA, Shanghai Jiao Tong University, Shanghai 200240, People’s Republic of China}

\author[]{David A. Velasco-Romero}
\email{david.velasco@princeton.edu}
\affiliation{School of Natural Sciences, Institute for Advanced Study, NJ 08540, USA}
\affiliation{Department of Astrophysical Sciences, Princeton University, 4 Ivy Lane, Princeton, NJ 08544, USA}

\author[0009-0004-6973-3955]{Fangyuan Yu}
\email{fy6204@princeton.edu}
\affiliation{Department of Astrophysical Sciences, Princeton University, 4 Ivy Lane, Princeton, NJ 08544, USA}

\author[orcid=0009-0004-8669-2411]{Jing-Ze Xia}
\email{jackxia@sjtu.edu.cn}
\affiliation{Tsung-Dao Lee Institute, Shanghai Jiao Tong University, Shengrong Road 520, Shanghai, 201210, China}

\author[0000-0003-0750-3543]{Xinyu Li}
\email{xinyuli@tsinghua.edu.cn}
\affiliation{Department of Astronomy, Tsinghua University, 30 Shuangqing Rd, Beijing, 100084, China}

\author[orcid=0000-0002-8131-6730]{Yosuke Mizuno}
\email{mizuno@sjtu.edu.cn}
\affiliation{Tsung-Dao Lee Institute, Shanghai Jiao Tong University, Shengrong Road 520, Shanghai, 201210, China}
\affiliation{School of Physics and Astronomy, Shanghai Jiao Tong University, 
800 Dongchuan Road, Shanghai, 200240, China}
\affiliation{Key Laboratory for Particle Physics, Astrophysics and Cosmology, Shanghai Key Laboratory for Particle Physics and Cosmology, Shanghai Jiao-Tong University, 800 Dongchuan Road, Shanghai, 200240, China}

\correspondingauthor{Hong-Xuan Jiang}
\email{hongxuan\_jiang@sjtu.edu.cn}
\begin{abstract}
In a tidal disruption event (TDE), a black hole (BH) tears apart a star, whose bound debris returns over many orbits to form an accretion disk. We present a framework built on the GPU-accelerated finite-volume code \texttt{AthenaK} that follows these stages in a single simulation with self-gravity throughout. We extend \texttt{AthenaK} with a moving frame, restart remapping between domains, an adaptive-mesh multigrid Poisson solver, a tabulated hydrogen--helium equation of state including recombination, a dual-energy update for the cold supersonic debris stream, a moving BH potential with excision, and localized adaptive time stepping (LAT), which speeds up the production run more than threefold. Each component is validated separately and in combination: the gravity solver maintains an isolated Lane--Emden sphere to a density error of $2.0\times10^{-3}$ over 16.3 dynamical times, and dual-energy recovery reduces the pressure error of a Mach $7.75\times10^7$ entropy wave from 15.7 to $4.0\times10^{-12}$. We demonstrate the framework with a Newtonian $\beta=1$ disruption of a $1\,M_\odot$, $1\,R_\odot$ star by a $10^3\,M_\odot$ BH. The debris has the expected energy spread, insensitive to the self-gravity update interval, splits evenly into bound and unbound material, and yields a fallback rate approaching $t^{-5/3}$ at late times. At the pericenter nozzle, the thermal energy gained in the high-resolution run matches the vertical kinetic energy lost to within 6\%, whereas the fiducial run, with 4--8 times coarser cells, overheats the thinnest early stream through numerical dissipation. Their heating agrees to 10\% once the returning stream thickens. Multifrequency LTE post-processing produces synthetic images and luminosities.

\end{abstract}



\section{Introduction}
When a star passes close to a black hole (BH), the tidal force from the BH can exceed the self-gravity of the star.
The star is converted into an extended stream with a broad distribution of orbital energies. 
Part of the stellar debris returns to the BH and is compressed and shocked before forming an accretion disk.
This catastrophic phenomenon is called a tidal disruption event (TDE).
Numerical simulations are essential to understand the dynamic process of TDEs and pose a numerical challenge to computational astrophysics.
The evolution is intrinsically a three-dimensional, multi-scale, multi-physics problem that couples gravity, compressible hydrodynamics, and thermodynamics with rapidly changing geometry \citep{2021ARA&A..59...21G,2021SSRv..217...40R,2020SSRv..216...63L}. 

Previous studies usually focus on different stages of the evolution and complement each other with different numerical approaches.
Early Eulerian calculations with adaptive mesh refinement (AMR) established the big picture of how stellar debris falls back to the BH and the energy disruption for various stellar structures and penetration factors \citep{1989ApJ...346L..13E,2013ApJ...767...25G}. Lagrangian smoothed particle hydrodynamics (SPH) provided better calculations for the expanding stream with high resolution, where the mass resides. 
However, its treatment of shocks relies on artificial viscosity and conductivity, whose numerical effects 
and convergence properties must be carefully assessed \citep{2016MNRAS.455.2253B,2017A&A...600A.124M,2019MNRAS.485..819L,2024ApJ...971L..46P}. 
Fixed-grid Godunov schemes provide conservative finite-volume shock capturing and direct AMR control. 
However, solving the fluid equations in a
large low-density regions with gas localized in the stellar debris makes them computationally inefficient. 
Moving-mesh methods achieve quasi-Lagrangian advection with Godunov-type fluxes \citep{2019MNRAS.487..981G,2024Natur.625..463S}, though their shock capturing can suffer from mesh distortion and typically requires additional regularization compared to fixed-grid schemes. 
Hybrid computations that follow the disruption with SPH and then map the debris onto grids in Newtonian, general-relativistic hydrodynamics (GRHD) or general-relativistic magnetohydrodynamics (GRMHD) combine realistic stellar initial conditions with grid-based shock capturing. However, these treatments introduce interpolation errors and start-up transients that must be explicitly checked. 
Grid-based calculations that follow the returning stream to pericenter either omit debris self-gravity or retain it only while the star is compact, and switch it off once the debris is mapped onto a global grid \citep{2020ApJ...904...99R,2023ApJ...957...12R,2026ApJ..1003..221C}. This can underestimate the transverse confinement and compactness of the stream \citep{1994ApJ...422..508K,2015ApJ...808L..11C,2016MNRAS.455.3612C}. 
Spherical grids have high resolution near the BH, but much lower resolution for the extended large-scale stream \citep{2015ApJ...804...85S,2016MNRAS.458.4250S,2020ApJ...904...99R,2022MNRAS.510.1627A,2023ApJ...957...12R,2026ApJ..1003..221C}. 
The hydrostatic star, dilute outbound debris, narrow returning stream, and shocked inner flow impose different requirements on resolution, conservation, and thermodynamic closure.

\begin{table*}
\centering
\caption{Numerical setup of the two realistic TDE simulations. The mass, length, and time scales listed in the first row define the code units. Both runs share the same initial condition and remapping sequence up to $t=0.79\,P_{\rm mb}$; table entries spanning both columns apply to both runs. The simulations differ only in the AMR configuration adopted during the fixed-BH fallback stage and in the duration over which this stage is evolved. The individual remap domains and measured debris masses are summarized in Table~\ref{tab:conservation-audits}. Here $P_{\rm mb}=2\pi GM_\bullet(2\Delta{\cal E}_{\rm mb})^{-3/2}$ is the Keplerian orbital period of the most-bound debris, with $\Delta{\cal E}_{\rm mb}=GM_\bullet R_*/r_{\rm t}^2$ (Eq.~\ref{eq:tde-most-bound-energy}). The code time unit is $t_0$, and $t$ is measured from the start of the simulation. The dimensionless thresholds $\eta_1$ and $\eta_2$ control the dual-energy pressure-recovery switch (Eq.~\ref{eq:dual-primitive-switch}) and the auxiliary-energy synchronization (Eq.~\ref{eq:dual-sync}), respectively. The interval between successive self-gravity Poisson solves is $\Delta t_{\rm sg}$. The softening length of the BH potential (Eq.~\ref{eq:bh-softened-gravity}) is $r_{\rm soft}$, and the radius of the excised sink region is $r_{\rm exc}$. The number of quadrature sample points per target cell in the remap operator (Eq.~\ref{eq:remap-cell-average}) is $N_{\rm qd}$, and $f_{\max}$ is the largest power-of-two timestep factor that LAT may assign to a MeshBlock (Eq.~\ref{eq:lat-factor}).}
\label{tab:fiducial-numerical-setup}
\small
\begin{tabular}{lll}
\hline\hline
Category & \texttt{FID} & \texttt{HR} \\
\hline
Code units & \multicolumn{2}{l}{$M_0=1\,M_\odot$, $L_0=2\,R_\odot$ ($R_*=0.5$); $t_0=1.2709\times10^3\,\mathrm{s}$} \\
 & \multicolumn{2}{l}{$v_0=1.0948\times10^8\,\mathrm{cm\,s^{-1}}$; $\rho_0=0.7382\,\mathrm{g\,cm^{-3}}$} \\
TDE model & \multicolumn{2}{l}{$M_*=1\,M_\odot$, $R_*=1\,R_\odot$, $M_\bullet=1000\,M_\odot$} \\
 & \multicolumn{2}{l}{$\beta=1$, $e_{\rm orb}=1$, $r_{\rm t}=r_{\rm p}\simeq10\,R_\odot$, $P_{\rm mb}=1.295\,\mathrm{d}=88.04\,t_0$} \\
Orbit/frame & \multicolumn{2}{l}{Initial separation $10\,r_{\rm t}=100\,R_\odot$; translating frame $\rightarrow$ fixed-BH frame} \\
Shared remaps & \multicolumn{2}{l}{$64^3\rightarrow256^3\rightarrow512^2\times256\,R_\odot^3$ for $t\leq0.79\,P_{\rm mb}$; MeshBlocks of $32^3$ zones} \\
Hydrodynamics & \multicolumn{2}{l}{RK2, PLM \citep{1979JCoPh..32..101V}, HLLE \citep{1988SJNA...25..294E}} \\
 & \multicolumn{2}{l}{$\mathrm{CFL}=0.4$ in the initial $64^3$ box and $0.5$ after the first remap} \\
 & \multicolumn{2}{l}{Density/pressure floors $\rho_{\rm floor}=10^{-10}\rho_0$, $P_{\rm floor}=10^{-12}\rho_0v_0^2$} \\
EOS & \multicolumn{2}{l}{Chabrier--Tomida--HELM H/He table; $X_{\rm H}=0.7$, $Y_{\rm He}=0.3$; H$_2$ and radiation pressure} \\
Dual energy & \multicolumn{2}{l}{$\eta_1=10^{-3}$, $\eta_2=10^{-4}$} \\
Gravity/BH & \multicolumn{2}{l}{Self-gravity cadence $\Delta t_{\rm sg}=0.03\,t_0$; softened Newtonian BH potential} \\
 & \multicolumn{2}{l}{$r_{\rm soft}=1\times10^{-2}\,R_\odot$; excision radius $r_{\rm exc}=0.4\,R_\odot$} \\
Remap operator & \multicolumn{2}{l}{Quadrature-sampled conserved-variable remap, $N_{\rm qd}=2^3$ subcell samples} \\
\hline
\multicolumn{3}{l}{\emph{Fixed-BH fallback stage}} \\
Refinement levels & {8} & 10 \\
Finest cells & {$\Delta x=\Delta z=6.25\times10^{-2}\,R_\odot$} & $\Delta x=1.5625\times10^{-2}$, $\Delta z=7.8125\times10^{-3}\,R_\odot$ \\
$r_{\rm t}/\Delta x$, $r_{\rm t}/\Delta z$ & {160, 160} & 640, 1280 \\
LAT & {7 LAT levels; $f_{\max}=128$} & 7 LAT levels; $f_{\max}=128$ \\
End time & {$5.00\,P_{\rm mb}$} & $2.94\,P_{\rm mb}$ \\
\hline
\multicolumn{3}{l}{Synthetic observables: AMR-to-ray post-processing; EOS temperature; multifrequency continuum transfer} \\
\hline
\end{tabular}
\end{table*}

Numerical computation of the full TDE process in a single simulation is a great challenge.
The spatial size of the debris expands several orders of magnitude compared to the initial star, while the returning stream and pericenter layers remain geometrically thin, which require extremely high resolution. 
A fixed domain large enough for the late flow is inefficient during the early disruption stage, whereas a compact domain cannot contain the expanding debris.  
Calculations have shown that the debris can stay eccentric and dissipate energy through several distinct interactions \citep{2013MNRAS.434..909H,2015ApJ...804...85S,2020MNRAS.495.1374B,2024Natur.625..463S,2024ApJ...971L..46P}. 
The high spatial resolution required to resolve the shock produces extremely small Courant timesteps near the BH, making long-duration simulations prohibitively expensive.
The gas in the debris passes through molecular, ionization, and radiation-pressure regimes in which latent heat and the varying mean molecular weight become important \citep{2026OJAp....9E.106A, 2024Natur.625..463S}. 
An efficient Eulerian workflow must therefore move, enlarge, and reorient the mesh while preserving orbital state, conservation, and thermodynamic consistency, as well as mitigating the severe timestep penalty associated with high enough resolution for the nozzle shock near the BH.

The flow near the pericenter provides a particularly demanding test of numerical fidelity. 
The stream will collapse to an extremely thin nozzle, while recombination and H$_2$ formation modify its incoming width \citep{2022MNRAS.511.2147B,2024Natur.625..463S,2026OJAp....9E.106A}. Higher-resolution SPH calculations find sharply decreasing pre-intersection spreading and dissipation \citep{2026ApJ...996L..21H,2026ApJ...999L..40K}. 
Moreover, the transition from longitudinal divergence to convergence can trigger spurious heating through either SPH viscosity switches or Godunov Riemann solvers \citep{2026arXiv260520327N}. 
Measured heating must therefore be diagnosed directly and not inferred from morphology; the choice of particle or grid representation alone does not establish physical convergence.

In this paper, we present an end-to-end TDE framework built on the performance-portable \texttt{AthenaK} finite-volume code \citep{2026ApJS..283...27S}.\footnote{The code, the example input files, and the radiative post-processing package are publicly available at \url{https://github.com/HongxuanJiang/athenak-transients}.} 
We add and couple a translating frame, a moving analytic BH potential and excision boundary, AMR gas self-gravity, a composite hydrogen/helium equation of state (EOS), domain remapping and Galilean frame conversion, dual-energy pressure recovery, and localized adaptive time stepping (LAT) following \citep{2022ApJS..263...26L}.  
The contribution lies in the coupled framework rather than in any single component. Together these elements allow a single calculation to follow the star from hydrostatic equilibrium through disruption to fallback, whereas grid-based studies have generally followed the disruption and the fallback in separate calculations connected by a mapping \citep[e.g.,][]{2023ApJ...957...12R,2026ApJ..1003..221C}. The physical interpretation of this calculation is presented separately \citep{Jiang2026Science}.
We validate this framework using controlled module tests with an end-to-end calculation of a $1\,M_\odot$, $1\,R_\odot$ star disrupted by a BH with a mass of $10^3\,M_\odot$, focusing on key debris, fallback, and nozzle-shock diagnostics. 
Local thermal equilibrium (LTE) radiative post-processing produces synthetic images and instantaneous luminosities. While a full, systematic convergence study would be computationally prohibitive and is beyond the scope of this work, comparing the nozzle-shock energy budget between the two production resolutions offers valuable insight into numerical fidelity and physical consistency.

The paper is organized as follows.  Sec.~\ref{sec:overview} introduces the physical setup and computational workflow. Secs.~\ref{sec:moving-frame}, \ref{sec:eos}, \ref{sec:remap}, \ref{sec:dual-energy}, and \ref{sec:lat} describe the translating-frame and self-gravity treatment, equation of state, domain remapping, dual-energy formulation, and localized adaptive time-stepping, respectively.  Sec.~\ref{sec:integrated-tde-validation} presents the end-to-end validation, Sec.~\ref{sec:radiative-postprocessing} describes the radiative post-processing, and Sec.~\ref{sec:summary-limitations} summarizes the main results and limitations.

\begin{figure*}
  \centering
  \includegraphics[width=\textwidth]{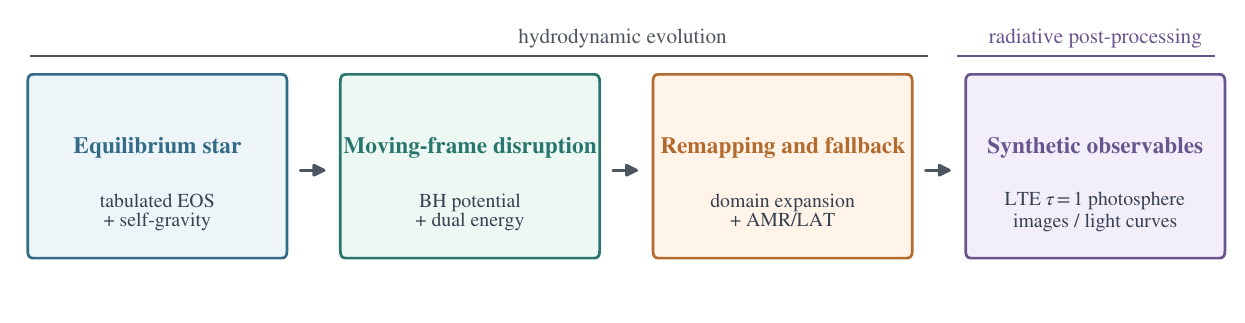}
  \caption{Compact workflow of the coupled TDE calculation. An EOS-balanced stellar model is disrupted in a moving frame, followed through domain remapping and fixed-BH fallback, and finally mapped to view-dependent synthetic observables. The principal numerical ingredients associated with each stage are listed inside the boxes.}
  \label{fig:tde-workflow}
\end{figure*}

\section{Physical setup and computational workflow}\label{sec:overview}

The hydrodynamic solver, standard AMR machinery, and task infrastructure are inherited from the public \texttt{AthenaK} code \citep{2026ApJS..283...27S}. 
The translating frame, self-gravity coupling\footnote{During the preparation of this manuscript, a self-gravity module was released in the public \texttt{AthenaK} repository. 
The implementation used here was developed from an earlier preliminary version and subsequently restructured and optimized for the TDE workflow. 
It therefore differs in implementation details from the current public module.}, 
EOS construction, remapping, dual-energy update, BH excision, and LAT implementation described below are the TDE-specific extensions tested in this work.

A TDE simulation is naturally divided into two stages. In the first stage, a star of mass $M_*$ and radius $R_*$ approaches a BH of mass $M_\bullet$ on an orbit with pericenter distance $r_{\rm p}$ comparable to or smaller than the tidal radius,
\begin{equation}
  r_{\rm t}=R_*\left(\frac{M_\bullet}{M_*}\right)^{1/3} .
  \label{eq:tidal-radius}
\end{equation}
During this encounter, the stellar structure and the onset of disruption must be accurately resolved. We therefore begin in a non-rotating frame that translates with the star or with the dense debris. In this frame, a compact, high-resolution computational domain remains centered on the disrupted material, while the BH appears as a moving external potential.
As the debris stream expands, the simulation is repeatedly remapped onto larger computational domains. The BH is treated as a live sink particle, so its position and velocity are updated throughout the calculation, and the translating-frame state is adjusted consistently. After the star is disrupted and the residual BH drift through the translating moving-frame coordinates has become negligible, the conserved state is Galilean transformed to the BH inertial frame and the BH position is fixed in the BH inertial coordinates. 
Later domain enlargements, AMR retargeting, and LAT fallback evolution are then performed in this BH inertial frame.

Throughout this paper, we focus on one fiducial Newtonian TDE model, a $1\,M_\odot$, $1\,R_\odot$ star disrupted by a $M_\bullet=1000\,M_\odot$ black hole. The stellar model is initialized in hydrostatic equilibrium with the tabulated EOS (hydrogen and helium mass fractions $X_{\rm H}=0.7$ and $Y_{\rm He}=0.3$). The orbit is parabolic, with orbital eccentricity $e_{\rm orb}=1$ and penetration factor $\beta\equiv r_{\rm t}/r_{\rm p}=1$, so that $r_{\rm p}=r_{\rm t}\simeq10\,R_\odot$ for this mass ratio. The star starts at a separation of $10\,r_{\rm t}$ and passes pericenter at $t\simeq0.24\,P_{\rm mb}$, where $P_{\rm mb}\simeq1.3$~d is the orbital period of the most-bound debris (Table~\ref{tab:fiducial-numerical-setup}). This setup is evolved as two production calculations that share the same initial condition, the same physics, and the same remapping sequence up to $t=0.79\,P_{\rm mb}$, and that differ only in the AMR configuration adopted for the fixed-BH fallback stage. We label them \texttt{FID} and \texttt{HR} (Table~\ref{tab:fiducial-numerical-setup}). \texttt{FID} continues on the resolution reached at the last shared remapping, $\Delta x=\Delta z=6.25\times10^{-2}\,R_\odot$, so that $r_{\rm t}$ is covered by $160$ cells in every direction, and is carried up to $5.00\,P_{\rm mb}$.  \texttt{HR} is remapped once more at $t=0.79\,P_{\rm mb}$ onto an AMR hierarchy whose finest-level cell spacing is 4 times smaller in the orbital plane and 8 times smaller vertically ($r_{\rm t}/\Delta x=640$, $r_{\rm t}/\Delta z=1280$), providing a much higher resolution for the nozzle and stopped at $2.94\,P_{\rm mb}$.  
The choice $M_\bullet=1000\,M_\odot$ targets an intermediate-mass-BH TDE \citep{2018NatAs...2..656L,2020ARA&A..58..257G} and permits several fallback times to be followed at high spatial resolution. The framework is not intrinsically tied to this mass ratio, but every production result in this paper uses the softened Newtonian potential in Eq.~(\ref{eq:bh-softened-gravity}); pseudo-Newtonian \citep{2013MNRAS.433.1930T} or relativistic gravity \citep{2020ApJ...904..101R,2026ApJ..1003..221C} and the longer fallback times associated with larger BH masses will be investigated in our future work. A relativistic calculation with a pericenter of about ten gravitational radii finds that the debris still remains highly eccentric \citep{2026ApJ..1003..221C}.

Fig.~\ref{fig:tde-workflow} summarizes the four stages of the calculation, Fig.~\ref{fig:remap-density-sequence} follows the production calculation across the successive domains (Sec.~\ref{sec:remap}), and Fig.~\ref{fig:tde-rt-multiview} shows a late snapshot together with its synthetic images (Sec.~\ref{sec:radiative-postprocessing}).

\section{Moving-frame dynamics and self-gravity}\label{sec:moving-frame}
\subsection{Moving-frame dynamics and BH treatment}

\begin{figure*}
  \centering
  \includegraphics[width=.75\textwidth]{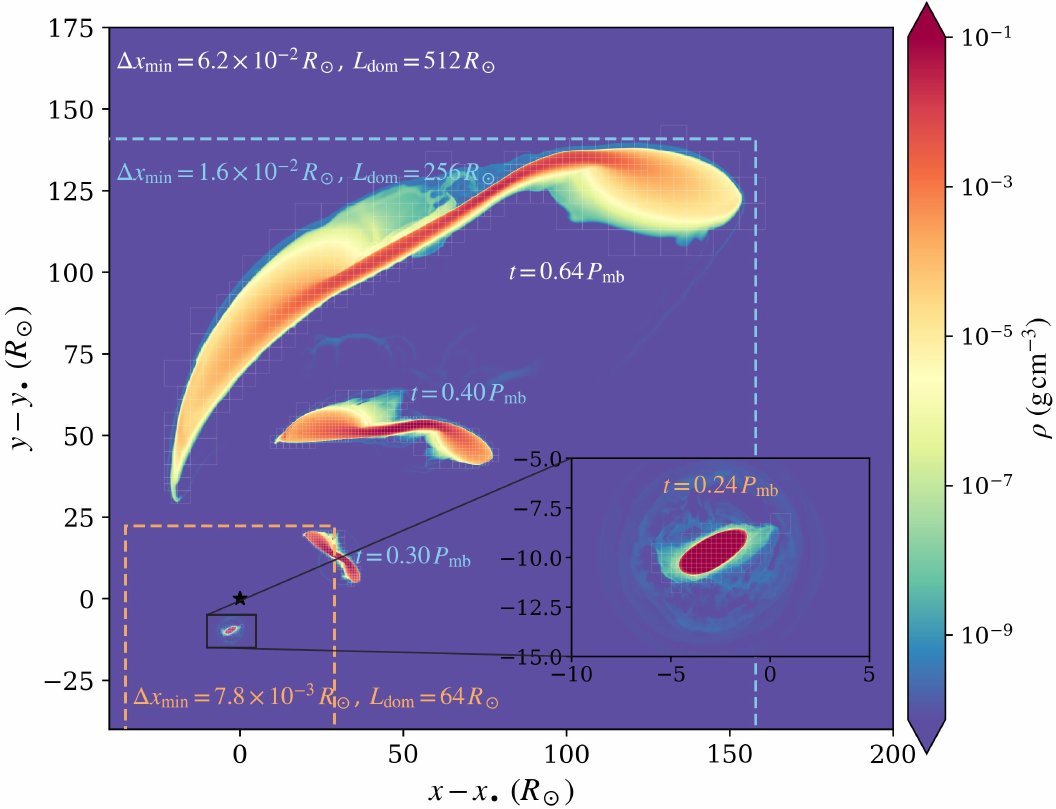}
  \caption{BH-centered midplane density slices during the remapping sequence. The density is shown in cgs units, and the black star marks the BH position. Dashed boxes show the domains used at the selected stages; the outline at $t/P_{\rm mb}=0.40$ is omitted for visual clarity. Thin gray lines mark MeshBlock boundaries where the density exceeds $10\rho_{\rm floor}$. The inset zooms in on the first remap around the star and BH. A movie of the midplane density in the \texttt{FID} run is available at \url{https://youtu.be/Roow6KcmfUI}.
  }
  \label{fig:remap-density-sequence}
\end{figure*}

\begin{figure*}
  \centering
  \includegraphics[width=\textwidth]{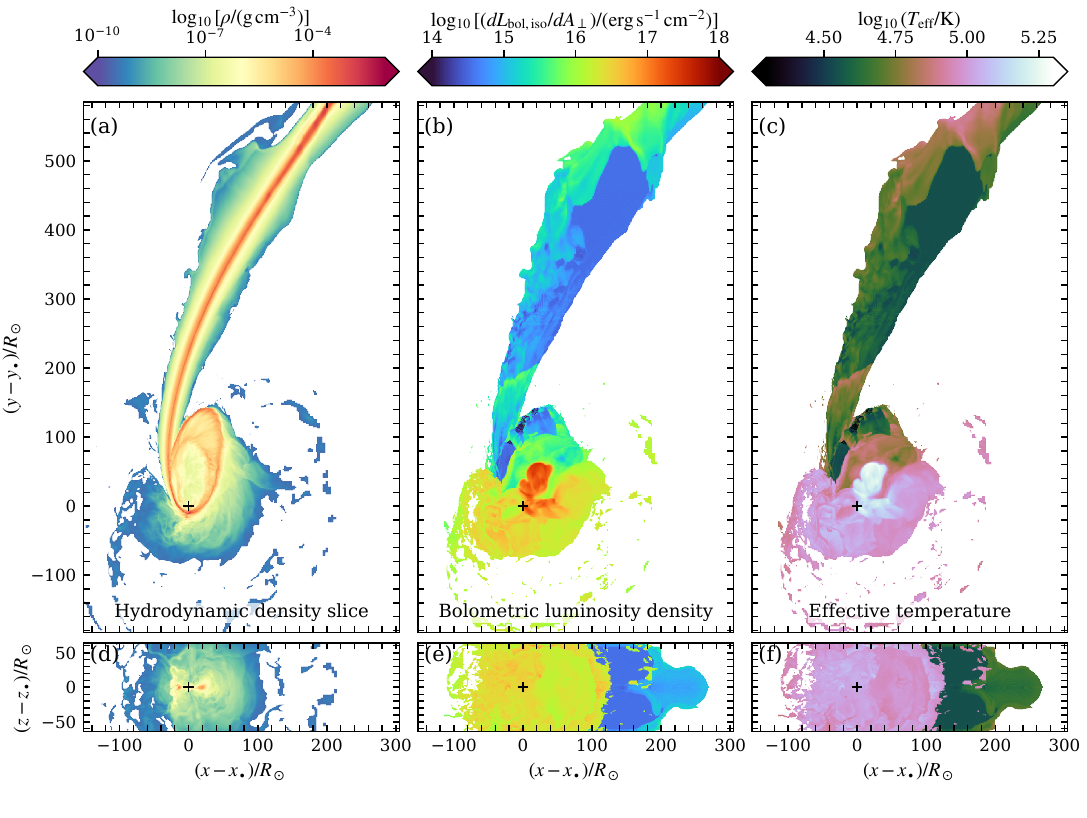}
\caption{Matched hydrodynamic and LTE radiative post-processing views of the last \texttt{HR} snapshot at $t/P_{\rm mb}=2.93$, rendered on $1024^2$ image-plane pixels with 1024 samples per line of sight. Panels (a)--(c) show the face-on view from $+z$ toward $-z$, and panels (d)--(f) the in-plane view from $-y$ toward $+y$. Panels (a) and (d) are BH-centered density slices in the corresponding image planes. Panels (b) and (e) show the projected isotropic-equivalent bolometric luminosity density, $dL_{\rm bol,iso}/dA_\perp=4\pi I$, from the multifrequency continuum transfer, where $I$ is the frequency-integrated emergent intensity. Panels (c) and (f) show the corresponding effective temperature $T_{\rm eff}=(\pi I/\sigma_{\rm SB})^{1/4}$. Crosses mark the BH position. Integrating the two image planes gives $L_{\rm bol,iso}^{(+z)}=4.8\times10^{42}\,\mathrm{erg\,s^{-1}}$ and $L_{\rm bol,iso}^{(-y)}=2.9\times10^{42}\,\mathrm{erg\,s^{-1}}$ (see Sec.~\ref{sec:radiative-postprocessing}).}
  \label{fig:tde-rt-multiview}
\end{figure*}

Following the star in a translating frame is established in Newtonian disruption calculations \citep{2009ApJ...705..844G,2013ApJ...767...25G} and in relativistic ones, where the frame follows the star through the curved spacetime of the BH \citep{2013PhRvD..87j4010C,2020ApJ...904...99R,2026ApJ..1003..221C}. Here the frame is retained through repeated remaps and later converted to the BH inertial frame (Sec.~\ref{sec:remap}).
We use ${\cal E}_{\rm tot}$ and ${\cal U}$ for total- and internal-energy densities, $\varepsilon_{\rm th}={\cal U}/\rho$ for specific thermal energy (including the radiation, ionization, and dissociation contributions of the EOS), and ${\cal E}_{\rm orb}$ for BH-relative specific orbital energy.
Let ${\bf R}$ denote an inertial position and ${\bf v}_{\rm I}=d{\bf R}/dt$ the inertial gas velocity. The moving simulation frame has origin ${\bf R}_{\rm fr}(t)$, velocity ${\bf V}_{\rm fr}=d{\bf R}_{\rm fr}/dt$, and acceleration ${\bf A}_{\rm fr}=d{\bf V}_{\rm fr}/dt$.  The moving-frame coordinates and frame velocity are
\begin{equation}
  {\bf r}={\bf R}-{\bf R}_{\rm fr}(t),
  \qquad
  {\bf v}={\bf v}_{\rm I}-{\bf V}_{\rm fr}(t).
  \label{eq:frame-transform}
\end{equation}
Because the simulation frame moves but does not rotate, the inertial force is the spatially uniform acceleration $-{\bf A}_{\rm fr}$ only.  In this frame, the hydrodynamic equations are as follows
\begin{align}
  \frac{\partial \rho}{\partial t}+\nabla\cdot(\rho{\bf v}) &=0,\\
  \frac{\partial(\rho{\bf v})}{\partial t}
  +\nabla\cdot(\rho{\bf v}{\bf v}+P{\bf I})
  &=-\rho\nabla\Phi-\rho{\bf A}_{\rm fr},
  \label{eq:frame-momentum}\\
  \frac{\partial {\cal E}_{\rm tot}}{\partial t}
  +\nabla\cdot\left[({\cal E}_{\rm tot}+P){\bf v}\right]
  &={\cal S}_{\Phi,{\cal E}}+{\cal S}_{{\rm fr},{\cal E}},
  \label{eq:frame-energy}
\end{align}
where $\rho$ is the gas density, $P$ is the total (gas plus radiation) pressure returned by the EOS, and ${\bf I}$ is the identity tensor.
The gas total-energy density in the simulation frame is given as
\begin{equation}
  {\cal E}_{\rm tot}={\cal U}+\frac{1}{2}\rho |{\bf v}|^2.
\end{equation}
The gravitational potential is decomposed into an analytically-given BH contribution $\Phi_\bullet$ and a numerically-calculated self-gravity contribution $\Phi_{\rm sg}$,
\begin{equation}
  \Phi({\bf r},t)=\Phi_\bullet({\bf r},t)+\Phi_{\rm sg}({\bf r},t).
  \label{eq:potential-decomposition}
\end{equation}
The prescribed BH contribution is represented by the softened Newtonian potential
\begin{equation}
  \Phi_\bullet({\bf r},t)
  =-\frac{G M_\bullet}
  {\sqrt{|{\bf r}-{\bf r}_\bullet(t)|^2+r_{\rm soft}^2}} .
  \label{eq:bh-softened-gravity}
\end{equation}
We adopt $r_{\rm soft}=10^{-2}\,R_\odot$ which is well inside the excision radius $r_{\rm exc}=0.4\,R_\odot$. The excision boundary is a one-way sink: only inward fluxes from the outside BH cells are accepted, its interior is reset to the zero-momentum floor state after each step, and removed quantities are excluded from the analysis.
The source terms ${\cal S}_{\Phi,{\cal E}}$ and ${\cal S}_{{\rm fr},{\cal E}}$ denote the gravitational and frame-acceleration work.  In the implementation, the gravitational work is discretized with the same mass fluxes used by the finite-volume hydrodynamical update, as described below.

The moving frame is chosen to follow the stellar structure and disrupted debris stream. We estimate its acceleration from density-selected, non-excised gas as a mass-weighted BH acceleration,
\begin{equation}
  {\bf A}_{\rm fr}
  =
  \frac{\int_{\cal D}\rho({\bf r})\,{\bf g}_\bullet({\bf r})\,dV}
       {\int_{\cal D}\rho({\bf r})\,dV},
  \qquad
  {\cal D}=\left\{\rho>\rho_{\rm fr,min},\ {\bf r}\notin{\cal B}_\bullet\right\},
  \label{eq:frame-acceleration}
\end{equation}
where ${\bf g}_\bullet=-\nabla\Phi_\bullet$, $\rho_{\rm fr,min}$ is the density threshold used to exclude floor, and ${\cal B}_\bullet$ is the excised BH inner-boundary region. This choice keeps the frame tied to the star during the disruption and avoids spurious motion driven by floor-density gas.

Because $M_\bullet\gg M_*$, the BH is fixed in the inertial frame, with apparent moving-frame coordinates ${\bf r}_\bullet={\bf R}_\bullet-{\bf R}_{\rm fr}$ and ${\bf v}_\bullet=-{\bf V}_{\rm fr}$. Equation~(\ref{eq:frame-acceleration}) is evaluated in each hydrodynamic step, and a time-continuous BH trajectory supplies the potential and inertial source terms in intermediate Runge--Kutta (RK) time stages.

\subsection{Self-gravity solver}

Self-gravity is essential while the star is intact and remains important wherever the debris stream is sufficiently dense to stay compact and self-confined. We solve for the gas self-gravity potential from the Poisson equation
\begin{equation}
  \nabla^2 \Phi_{\rm sg} = 4\pi G \rho_{\rm sg},
  \label{eq:selfgrav-poisson}
\end{equation}
where $\rho_{\rm sg}$ is the gas density used as the source. 
In practice, we exclude floor-density material from the source term, so that the numerical atmosphere does not contribute to the long-range potential. 

Relativistic disruption calculations solve the Poisson equation on a uniform box with a spectral solver and image-mass boundary conditions and add the potential as a weak-field metric perturbation \citep{2013PhRvD..87j4010C,2020ApJ...904...99R,2026ApJ..1003..221C}. A multigrid solver on the adaptive mesh instead computes the potential on a deep refinement hierarchy and keeps it after the debris has expanded. We use a geometric multigrid method on the hydrodynamic AMR hierarchy, following the formulation developed for the \texttt{Athena++} AMR framework \citep{2023ApJS..266....7T}. Our implementation shares its origin with the public \texttt{AthenaK} solver (Sec.~\ref{sec:overview}) but differs in three respects. Its discrete operator accepts anisotropic cells, its hierarchy supports small root grids with deep refinement, and its potential is coupled to the localized time stepping of Sec.~\ref{sec:lat}.
We solve ${\cal A}_h\Phi_h=S_h$, where ${\cal A}_h\equiv-\nabla_h^2$ is a positive-definite, seven-point finite-difference operator and $S_h\equiv-4\pi G\rho_{{\rm sg},h}$. Same-level and coarse--fine ghost zones of $\Phi_{\rm sg}$ are refreshed before differentiation, ensuring a consistent gravitational acceleration across AMR interfaces. Multipole boundary conditions are supported for isolated configurations. 

Following standard geometric multigrid methods \citep{2000muto.book.....B}, we monitor the defect
\begin{equation}
  d_h = S_h - {\cal A}_h\Phi_h ,
  \label{eq:selfgrav-defect}
\end{equation}
through its volume-weighted root mean square over all leaf cells, $\|d_h\|_2=(V^{-1}\sum_i d_{h,i}^2\,\Delta V_i)^{1/2}$, where $\Delta V_i$ is the cell volume and $V$ the domain volume. The smoother is a red--black Gauss--Seidel update \citep{1996SJSC...17..180Y} with over-relaxation factor $\omega_{\rm GS}=1.15$ \citep{2023ApJS..266....7T}. For cell widths $(\Delta x,\Delta y,\Delta z)$, the neighbors in $y$ and $z$ enter the seven-point stencil with weights $w_y=(\Delta x/\Delta y)^2$ and $w_z=(\Delta x/\Delta z)^2$ and the central coefficient is $2(1+w_y+w_z)$, which reduces to the isotropic stencil for cubic cells. These weights are applied at every level, so the vertically refined \texttt{HR} domain with $\Delta z=\Delta x/2$ is solved with the same discrete operator as an isotropic mesh. In each V-cycle, the defect is volume-averaged over child cells and restricted to the next coarser level, the coarse-grid correction is computed in full-approximation form, and it is returned to the fine level by tricubic prolongation followed by post-smoothing. The hierarchy extends from the finest MeshBlocks through the parent cells of the refined levels into a root grid, which is coarsened down to a single cell and solved exactly as in \citet{2023ApJS..266....7T}. For periodic problems, the mean of the source term is subtracted in two passes, because a single pass leaves a rounding residual in the null space of the operator that limits the attainable convergence. Full multigrid is used for startup and remapped states, and all subsequent solves start from the previous potential and use ordinary V-cycles.

Solving Eq.~(\ref{eq:selfgrav-poisson}) at every Runge--Kutta stage would dominate the cost of the calculation. In a collapse test with five levels of adaptive refinement, the gravity solve takes 90\% of the wall time, and refreshing the potential every fourth cycle reduces the total by 79\%. Production runs therefore update the potential at a physical cadence $\Delta t_{\rm sg}$, starting each solve from the previous solution. For $\beta=1$, the stellar dynamical time $(R_*^3/GM_*)^{1/2}$ and the pericenter passage time $(r_{\rm p}^3/GM_\bullet)^{1/2}$ are both $1.25\,t_0$, so the production cadence $\Delta t_{\rm sg}=0.03\,t_0$ refreshes the potential about 40 times per dynamical time. Its effect on the disruption is tested in Sec.~\ref{sec:selfgrav-cadence}.

Table~\ref{tab:mg-cost} compares our solver with the public \texttt{AthenaK} solver on the same collapse test and hardware. One V-cycle costs about four times less in our implementation, and the whole test runs 4.4 times faster with the same converged solution. In this regime, a GPU V-cycle is limited by launch and communication latency rather than by arithmetic. We therefore launch kernels asynchronously, overlap halo exchanges with interior smoothing, and keep the coarse end of the hierarchy, a few hundred cells at most, on the host. Our V-cycle also post-smooths every intermediate level and ends with an extra sweep on the finest level, which reduces the number of V-cycles per solve.

\begin{table*}
\centering
\caption{Performance of the multigrid solver in the Bonnor--Ebert collapse test of the \texttt{AthenaK} test suite on two V100 GPUs, comparing this work with the public \texttt{AthenaK} solver run from the same input file. The isothermal sphere collapses in a box of $32^3$ code units on a root grid of $128^3$ or $256^3$ cells with five levels of adaptive refinement, over 100 steps. Both codes use two pre- and two post-smoothing sweeps, $\omega_{\rm GS}=1.15$, a warm start from the previous potential, and a defect tolerance of $10^{-9}$ at every stage. $\|d_h\|_2$ is the median final defect of Eq.~(\ref{eq:selfgrav-defect}) in code units. The time per V-cycle is the wall-time difference between runs with 6 and 2 fixed V-cycles per solve divided by the extra V-cycles. The time per solve is the wall time in excess of the same run without gravity divided by the number of solves.}
\label{tab:mg-cost}
\begin{tabular}{lcccccccc}
\hline
MeshBlock & \multicolumn{2}{c}{$\|d_h\|_2$ at convergence} & \multicolumn{2}{c}{V-cycles per solve} & \multicolumn{2}{c}{Time per V-cycle (ms)} & \multicolumn{2}{c}{Time per solve (ms)} \\
size & this work & public & this work & public & this work & public & this work & public \\
\hline
$16^3$ & $1.6\times10^{-10}$ & $3.6\times10^{-10}$ & 6.0 & 8.8 & 6.2 & 22.4 & 44 & 210 \\
$32^3$ & $2.0\times10^{-10}$ & $2.9\times10^{-10}$ & 5.1 & 6.3 & 11.3 & 49.6 & 68 & 356 \\
\hline
\end{tabular}
\end{table*}

When the self-gravity potential is obtained, the update from gravitational source term follows the formulation used for flux-consistent self-gravitating hydrodynamics in \texttt{Athena++}
\citep{2021ApJS..252...30M}. We use centered differences of the potential in
Eq.~(\ref{eq:potential-decomposition}) to update momentum, while the gravitational work is calculated in a flux form using the Godunov mass fluxes.
In one coordinate direction, the discrete update is
\begin{align}
  (\rho v_x)^{n+1}_{i}
  &=
  (\rho v_x)^{n}_{i}
  - \rho_i \Delta t\,
  \frac{\Phi_{i+1}-\Phi_{i-1}}{2\Delta x},
  \label{eq:lat-grav-momentum}\\
  {\cal E}_{{\rm tot},i}^{n+1}
  &=
  {\cal E}_{{\rm tot},i}^{n}
  - \frac{\Delta t}{2\Delta x}
  \Big[
  F_{\rho,i+1/2}(\Phi_{i+1}-\Phi_i)
  \nonumber\\
  &\qquad
  +
  F_{\rho,i-1/2}(\Phi_i-\Phi_{i-1})
  \Big].
  \label{eq:lat-grav-energy}
\end{align}
Equation~(\ref{eq:lat-grav-energy}) is the discrete gravitational work associated with the same mass flux used by the hydrodynamic update. This is important in a moving, cold debris stream because the total energy can be
dominated by kinetic and gravitational terms when the thermal energy is small.

The self-gravity potential is updated at the fixed times $k\,\Delta t_{\rm sg}$, so that a restarted calculation solves on the same cycles as an uninterrupted one. After regridding, the multigrid solver can be initialized with a valid potential transferred from the previous AMR hierarchy. If no such potential is available, a new self-gravity calculation is performed before gravity source terms are applied.

\subsection{Self-gravity validation}

We validate the self-gravity coupling with two tests. A linear Jeans wave provides an analytic time-dependent solution and tests the coupling between the Poisson solver and the hydrodynamic update. An isolated Lane--Emden polytrope tests that the solver maintains a nonlinear, self-gravitating hydrostatic equilibrium on the AMR mesh.

In the Jeans-wave test, we use an isothermal gas with a small density perturbation,
\begin{equation}
  \rho({\bf r},0)=\bar\rho\left[1+a_{\rm J}\cos({\bf k}\cdot{\bf r})\right],
  \qquad a_{\rm J}=10^{-3}.
  \label{eq:jeans-initial}
\end{equation}
The perturbation is initialized with zero velocity. For a stable Jeans mode,
$k>k_{\rm J}$, linear theory gives
\begin{equation}
  \delta\rho_k(t)=a_{\rm J}\bar\rho\cos(\omega_{\rm J}t),
  \qquad
  \omega_{\rm J}^2=c_s^2 k^2-4\pi G\bar\rho ,
  \label{eq:jeans-analytic}
\end{equation}
where $c_s$ is the isothermal sound speed and
$k_{\rm J}=(4\pi G\bar\rho/c_s^2)^{1/2}$. 
Here we use $\lambda/\lambda_{\rm J}=0.8$, giving an oscillatory rather than unstable mode. We compare a periodic uniform grid with $256^3$ cells against a static mesh refinement (SMR) hierarchy that has the same finest resolution and refines the central region $-0.5<r_x,r_y,r_z<0.5$. Both cases use the same domain $[-1,1]^3$ and $32^3$ cells per MeshBlock, so the SMR setup additionally exercises coarse--fine multigrid coupling.

\begin{figure*}
  \centering
  \includegraphics[width=\textwidth]{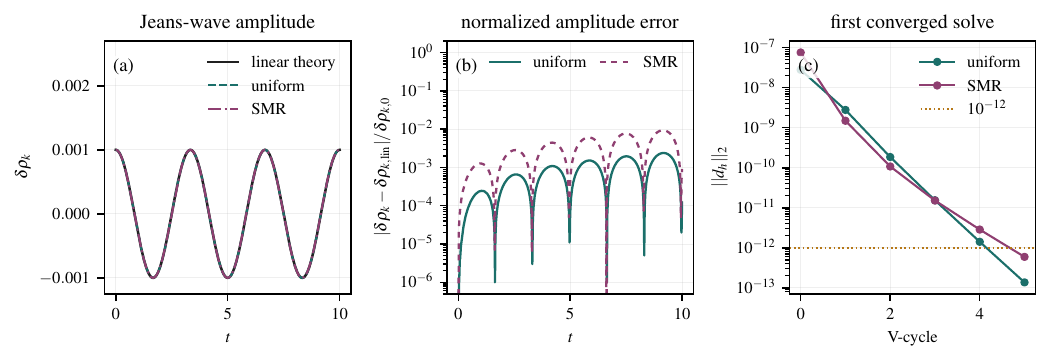}
  \caption{Validation of the multigrid self-gravity solver using a stable Jeans wave. Panel (a) compares the measured Fourier amplitude with the linear solution. Panel (b) shows the amplitude error normalized by the
  initial perturbation amplitude. {Panel (c) shows the multigrid defect as a function of V-cycle count for the first converged solve. V-cycle 0 is the full multigrid initial guess.}}
  \label{fig:selfgrav-jeans}
\end{figure*}

Fig.~\ref{fig:selfgrav-jeans} shows good agreement with the linear solution, with maximum normalized amplitude errors of $2.4\times10^{-3}$ on the uniform grid and $9.3\times10^{-3}$ on the SMR mesh, where the wave crosses refinement interfaces. Starting from the full multigrid initial guess, the first solve reaches the defect tolerance of $10^{-12}$ in five V-cycles on both meshes, with a contraction factor of about $0.1$ per V-cycle, consistent with \citet{2023ApJS..266....7T}.

\begin{figure}
  \centering
  \includegraphics[width=\columnwidth]{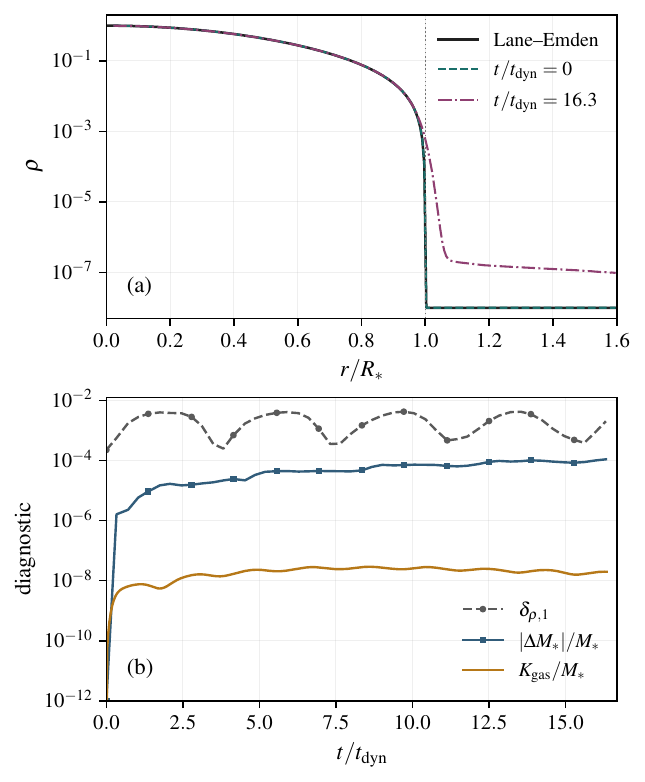}
  \caption{Nonlinear hydrostatic-equilibrium test using an isolated Lane--Emden polytrope. Panel (a) compares the initial and final spherically averaged density profiles with the analytic Lane--Emden profile. Panel (b) shows the radial profile error $\delta_{\rho,1}$, the mass drift $|\Delta M_*|/M_*$, and the specific kinetic energy $K_{\rm gas}/M_*$, where $K_{\rm gas}$ is the volume-integrated kinetic
  energy. Time is normalized by $t_{\rm dyn}=(R_*^3/GM_*)^{1/2}$.}
  \label{fig:selfgrav-polytrope}
\end{figure}

In the second test, we evolve an isolated polytropic star. The initial condition uses an $n=1.5$ Lane--Emden sphere with
$\gamma=1+1/n=5/3$, central density $\rho_c=1$, and radius $R_*=0.5$. 
The star is evolved in a computational domain of $[-3,3]^3$ with outflow boundaries. A fixed central refinement region at level $\ell=5$ covers the star, yielding a minimum cell size of $\Delta r_{\min}=5.86\times10^{-3}$ and about 170 cells across the stellar diameter. We measure time in units of
\begin{equation}
  t_{\rm dyn}=\left(\frac{R_*^3}{G M_*}\right)^{1/2},
  \label{eq:polytrope-tdyn}
\end{equation}
where $M_*$ is the Lane--Emden stellar mass.
The pressure profile follows from the polytropic relation and satisfies
\begin{equation}
  \nabla P = -\rho\nabla\Phi_{\rm sg}
  \label{eq:polytrope-equilibrium}
\end{equation}
in the continuum solution. Any secular contraction, expansion, or bulk motion therefore reflects the combined errors introduced by mapping the analytic profile to the Cartesian mesh, solving the Poisson equation, and applying the gravitational source term. We evolve the star with SMR and monitor its spherical density profile, stellar mass, and kinetic energy to track how these errors accumulate over time.
We quantify the profile drift using the radial-density $L_1$ error
\begin{equation}
  \delta_{\rho,1}(t)
  =
  \frac{
  \int_0^{R_*}
  \left|\langle\rho\rangle(r,t)-\rho_{\rm ref}(r)\right|\,dr
  }{
  \int_0^{R_*}
  \max\left[\rho_{\rm ref}(r)-\rho_{\rm floor},0\right]\,dr
  },
  \label{eq:stellar-profile-l1}
\end{equation}
where $\langle\rho\rangle$ is the spherical average of the numerical density. For the Lane--Emden test, $\rho_{\rm ref}$ is the analytic polytropic profile. For the tabulated-EOS test below, it is the initial EOS-balanced profile.

At $t/t_{\rm dyn}=16.3$, the star shows a density error of $2.0\times10^{-3}$, a mass drift of $1.1\times10^{-6}$, and a normalized kinetic energy $K_{\rm gas}/M_*<2.9\times10^{-8}$. These small residuals confirm that the nonlinear hydrostatic equilibrium remains stable over many dynamical times and validate the coupled performance of the stellar structure initialization, AMR hierarchy, and self-gravity solver.

\section{Equation of State}\label{sec:eos}

\subsection{Thermodynamic regimes in TDE}

As the TDE debris stream expands, its temperature can fall far below that of the initial star. Hydrogen and helium recombination and H$_2$ formation then change the pressure response and can modify the stream width and its subsequent interaction \citep{2024Natur.625..463S,2026OJAp....9E.106A}. A single gamma-law EOS cannot capture these transitions. During the course of the calculation, the gas passes through regimes that are molecular, partially ionized, fully ionized, dominated by radiation pressure, and heated by shocks. The cold stream may also retain only a small fraction of its total energy as its internal energy. Here, we use the tabulated-EOS interface in \texttt{AthenaK} and construct the composite H/He table. It is used in all processes in hydrodynamic primitive variable recovery, dual-energy pressure update, and stellar initialization.

\begin{figure*}
  \centering
  \includegraphics[width=.7\textwidth]{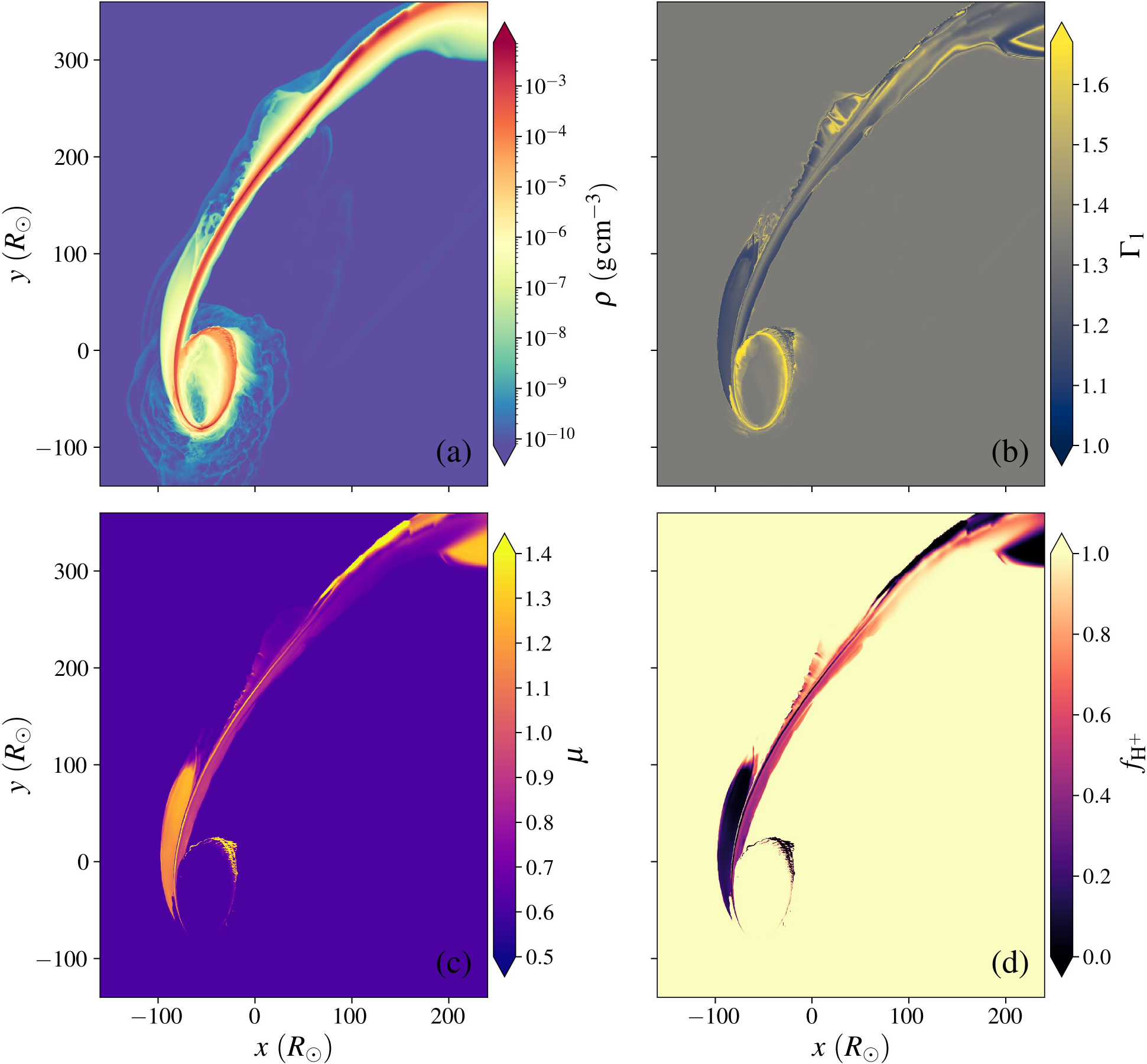}
  \caption{EOS diagnostics in a production TDE calculation at $t/P_{\rm mb}=1.70$. Panels (a)--(d) show the midplane density, first adiabatic exponent $\Gamma_1$, mean molecular weight $\mu$, and hydrogen ionization fraction $f_{\rm H^+}$. Distances are shown in solar
  radii, using $0.5$ code length units $=1\,R_\odot$. The strong spatial variation of these quantities shows that the debris occupies several thermodynamic regimes within a single, thermodynamically consistent EOS table.}
  \label{fig:eos-tde-diagnostics}
\end{figure*}

Fig.~\ref{fig:eos-tde-diagnostics} shows a representative slice at $z=0$ from run \texttt{HR}, with strong spatial variation in the first adiabatic exponent $\Gamma_1$, the mean molecular weight $\mu$, and the hydrogen ionization fraction $f_{\rm H^+}$. Expansion during outward motion and passage through apocenter experience cooling and recombination, while pericenter compression and the nozzle shock re-ionize the gas. A later collision between outgoing and incoming streams produces a distinct self-intersection shock, separate from the nozzle compression \citep{2024Natur.625..463S,2026OJAp....9E.106A}.

\subsection{Tabulated EOS construction}

For the tabulated EOS, we assume hydrogen and helium mass fractions $X_{\rm H}=0.7$ and $Y_{\rm He}=0.3$ with no metals. It is sampled on a $640\times640$ grid in
\begin{equation}
  \ln\rho_{\rm cgs},\,\,\mathrm{and}\,\, \ln T ,
\end{equation}
covering
\begin{equation}
  -20 \le \log_{10}\left(\frac{\rho}{\mathrm{g\,cm^{-3}}}\right) \le 3,
  \qquad
  0 \le \log_{10}\left(\frac{T}{\mathrm K}\right) \le 9 .
  \label{eq:eos-table-range}
\end{equation}
From the hydrodynamic evolution, these quantities are required:
\begin{equation}
  \ln P,\quad \ln\varepsilon_{\rm th},\quad \ln c_s^2,\quad
  \Gamma_1,\quad \Gamma_3-1 ,
  \label{eq:eos-table-fields}
\end{equation}
where $\varepsilon_{\rm th}$ is the specific thermal energy and $c_s$ is the adiabatic sound speed. The adiabatic derivatives are
\begin{equation}
  \Gamma_1 \equiv
  \left(\frac{\partial \ln P}{\partial \ln \rho}\right)_s,
  \qquad
  \Gamma_3-1 \equiv
  \left(\frac{\partial \ln T}{\partial \ln \rho}\right)_s ,
  \label{eq:eos-adiabatic-derivatives}
\end{equation}
so that $c_s^2=\Gamma_1 P/\rho$. The table also stores composition diagnostics, including the molecular hydrogen fraction $f_{\rm H_2}$, hydrogen ionization fraction $f_{\rm H^+}$, helium ionization fractions $f_{\rm He^+}$ and $f_{\rm He^{2+}}$, mean molecular weight $\mu$, and radiation pressure fraction $f_{\rm rad}\equiv P_{\rm rad}/P_{\rm tot}$. These quantities are interpolated for analysis and output, but they are not needed in the hydrodynamic evolution. 

\begin{figure*}
  \centering
  \includegraphics[width=\textwidth]{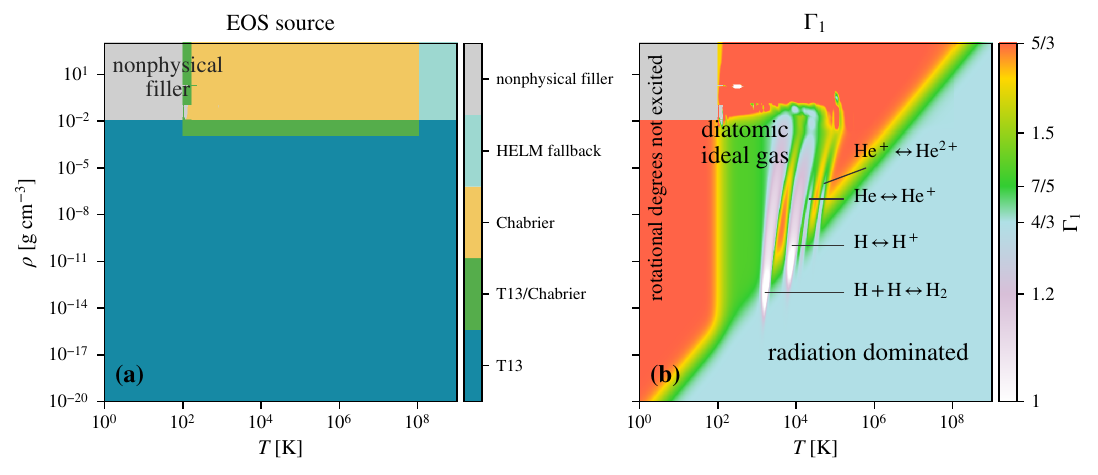}
\caption{EOS table employed in the TDE calculations. Panel (a) shows the constituent EOS model regions used to construct the grid. Here, ``T13'' refers to the ideal chemical equilibrium H/He EOS of \citet{2013ApJ...763....6T}, ``Chabrier'' refers to the dense fluid H/He EOS of \citet{2019ApJ...872...51C,2021ApJ...917....4C}, and ``T13/Chabrier'' indicates the transition region smoothly blended over the density range $10^{-3}$--$10^{-2}\,{\rm g\,cm^{-3}}$. At high temperatures exceeding the Chabrier table coverage, the fully ionized HELM style H/He branch \citep{2000ApJS..126..501T} serves as a fallback. Panel (b) shows the corresponding first adiabatic index $\Gamma_1$. The gray shaded region at low temperatures and high densities represents an unphysical filler appended solely to ensure a rectangular domain for robust 2D interpolation. This region is masked out and never sampled during the TDE simulations.}
\label{fig:eos-chabrier}
\end{figure*}

The table is assembled from three thermodynamic descriptions, as shown in Fig.~\ref{fig:eos-chabrier}. At low density, we use the ideal chemical-equilibrium H/He EOS of \citet{2013ApJ...763....6T}, which solves the dissociation and ionization equilibria of H$_2$, H, H$^+$, He, He$^+$, He$^{2+}$, and electrons together with charge neutrality. The equilibrium constants follow from the same partition functions as the thermodynamic energy, including the rotational and vibrational states of H$_2$ with a fixed ortho-to-para ratio of three. This branch therefore captures H$_2$ formation and recombination in the cold debris stream, along with the associated rapid changes in $\mu$, $\Gamma_1$, and $\Gamma_3-1$.

At high density and moderate temperature, we use the dense-fluid H/He EOS of \citet{2019ApJ...872...51C,2021ApJ...917....4C}, which replaces the ideal chemical branch where non-ideal effects are important. The transition between the Tomida and Chabrier branches is smoothed between $10^{-3}$ and $10^{-2}\ {\rm g\,cm^{-3}}$. At high temperatures where the dense Chabrier table does not provide coverage, we use a fully ionized HELM-style H/He fallback \citep{2000ApJS..126..501T}. In this part, hydrogen and helium are assumed fully ionized, while relativistic and degenerate $e^\pm$ thermodynamics are included in the same spirit as the HELM EOS. Photon radiation is then added as an independent blackbody component through a local LTE thermodynamic closure,
\begin{equation}
  P_{\rm rad}=\frac{a_{\rm rad}T^4}{3},
  \qquad
  \varepsilon_{\rm rad}=\frac{a_{\rm rad}T^4}{\rho},
  \label{eq:eos-radiation}
\end{equation}
and the stored pressure, internal energy, sound speed, and adiabatic derivatives are rebuilt for the combined EOS with radiation pressure.

At runtime, the EOS maps $(\rho, {\cal U}_{\rm EOS})$, where ${\cal U}_{\rm EOS}$ is the internal energy density (gas plus radiation) passed to the EOS closure and selected by the dual-energy switch of Eq.~(\ref{eq:dual-primitive-switch}), to the full thermodynamic state by interpolation within the tabulated domain, from which the pressure, sound speed, adiabatic derivatives, and composition diagnostics are all recovered.

\subsection{EOS-balanced stellar initialization}
\label{sec:eos-star}

The tabulated EOS is also used to construct the initial stellar model. For a gamma-law calculation, the code can initialize the star from a Lane--Emden polytrope. In a tabulated-EOS calculation, however, imposing a fixed polytropic pressure law and then switching closures would make the initial pressure support inconsistent with the EOS used during evolution. We therefore provide an EOS-balanced stellar initialization that constructs a one-dimensional spherical hydrostatic equilibrium profile with the EOS table by
integrating
\begin{align}
  \frac{d m_r}{dr} &= 4\pi r^2\rho,\\
  \frac{dP}{dr} &= -\frac{Gm_r\rho}{r^2},\\
  \frac{d\rho}{dr} &=
  \frac{\rho}{\Gamma_1(\rho,P)\,P}\,\frac{dP}{dr},
  \label{eq:eos-balanced-star}
\end{align}
where $\Gamma_1(\rho,P)$ is evaluated from the same EOS table used by the hydrodynamic solver. The central pressure is bracketed so that the pressure reaches zero at the requested stellar radius, and the central density is adjusted so that the integrated mass matches the requested stellar mass. The resulting profiles $\rho(r)$ and $P(r)$ are interpolated to the cell centers of the Cartesian mesh, and the internal-energy density is recovered from the same table through ${\cal U}=\rho\,\varepsilon_{\rm th}(\rho,P)$. Thus the initial pressure support, sound speed, and subsequent thermodynamic evolution all use a single EOS closure rather than a polytropic approximation patched onto a tabulated EOS.

\begin{figure}
  \centering
  \includegraphics[width=\columnwidth]{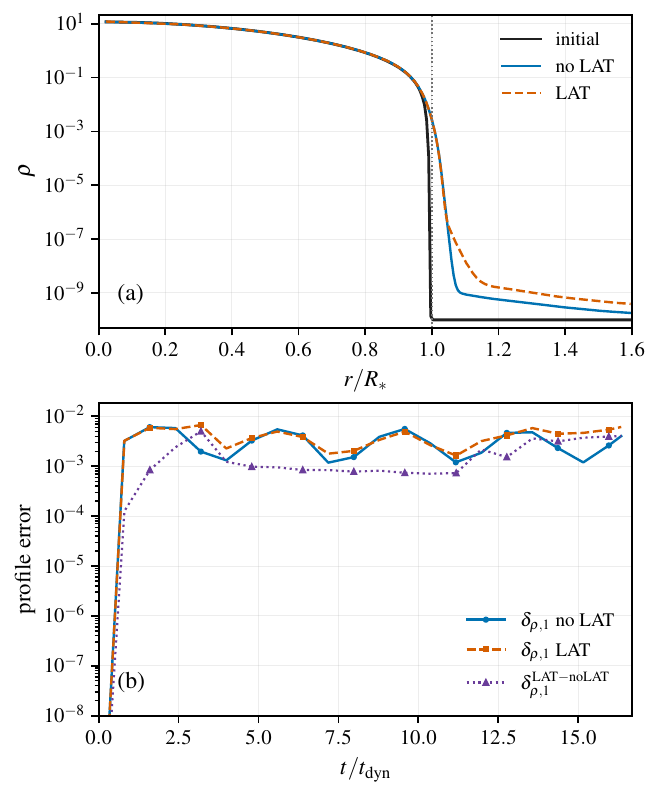}
  \caption{Long-duration LAT consistency test for the EOS-balanced star with self-gravity and the tabulated EOS. Panel (a) compares the initial
  density profile with the matched final profiles from the globally synchronized (no-LAT) and LAT runs. Panel (b) follows the profile error and the direct LAT--no-LAT difference through $t/t_{\rm dyn}=16.35$.}
  \label{fig:lat-eos-balanced-star}
\end{figure}

The initial profile closely matches an $n=1.5$ Lane--Emden model of the same mass and radius, indicating $\Gamma_1\simeq5/3$ throughout most of the interior. Evolved for $16.35\,t_{\rm dyn}$ with the production self-gravity cadence $\Delta t_{\rm sg}=0.03\,t_0$ (the globally synchronized run of Fig.~\ref{fig:lat-eos-balanced-star}), the star reaches a final profile error $\delta_{\rho,1}=4.0\times10^{-3}$ (maximum $6.1\times10^{-3}$), a mass drift of $3.5\times10^{-5}$, and $K_{\rm gas}/M_*<5.8\times10^{-6}$, so the tabulated-EOS model remains stable over many dynamical times without a closure mismatch.

\section{Domain remapping and frame conversion}
\label{sec:remap}

The debris grows from a star of one solar radius into a stream hundreds of solar radii long, while its densest parts stay geometrically thin, so no single fixed domain can resolve the whole evolution efficiently. At the beginning of a TDE simulation, we work in a moving frame that keeps the star near the center of a computational domain, allowing the disruption to be resolved without carrying the full late-time debris volume from the outset.
Once the debris stream has expanded beyond the initial high-resolution computational domain, we restart the calculation on a larger domain and remap the hydrodynamical variables from the old AMR hierarchy to the new one. The remapping is implemented as a restart operation: the old checkpoint supplies the source MeshBlock structure, hydrodynamic variables, physical time, and stored BH/frame metadata, while the new configuration defines the target computational domain, MeshBlock size, refinement limits, and AMR criteria.

Fig.~\ref{fig:remap-density-sequence} demonstrates the dynamical adjustment of the computational domain as the debris expands. Both \texttt{FID} and \texttt{HR} models undergo an identical sequence of domain enlargements during the early evolution. However, \texttt{HR} uses a final, targeted remap at $t=0.79\,P_{\rm mb}$ onto a domain with $768\times768\times128\,R_\odot^3$ to resolve the pericenter compression (Sec.~\ref{sec:nozzle}). Table~\ref{tab:conservation-audits} summarizes the resolution, mass, angular momentum, and bound fraction at different remapping stages shown in Fig.~\ref{fig:remap-density-sequence}. After subtracting the floor contribution, the debris mass remains within $1.1\times10^{-5}\,M_\odot$ of the initial $1\,M_\odot$, while the angular momentum changes relative to the BH by only $1.7\times10^{-3}$ over the archived sequence.

\begin{table*}
\centering
\caption{Conservation and state diagnostics along the remapping sequence. The frame label is read from each archived output header. All velocities used for $L_{{\rm I},z}$ are converted to the BH inertial frame. The debris mass is $M_{\rm deb}=\int\max(\rho-\rho_{\rm floor},0)\,dV$ outside the excision sphere. The mass row uses the initial $M_*$ as reference, whereas $L_{{\rm I},z}$ is normalized to its value at the first checkpoint. $N_{\rm MB}$ is the total number of MeshBlocks in the AMR hierarchy and $\Delta r_{\min}$ is the finest cell width. The axial angular momentum of the debris about the BH is $L_{{\rm I},z}=\int\rho\,[({\bf R}-{\bf R}_\bullet)\times({\bf v}_{\rm I}-{\bf v}_{\bullet,{\rm I}})]_z\,dV$, evaluated over the same volume as $M_{\rm deb}$. The bound fraction is $f_{\rm bound}=M_{\rm deb}({\cal E}_{\rm orb}<0)/M_{\rm deb}$, i.e., the mass fraction of the debris whose specific orbital energy relative to the BH, ${\cal E}_{\rm orb}$ of Eq.~(\ref{eq:tde-orbital-energy}), is negative. }
\label{tab:conservation-audits}
\begin{tabular}{lcccc}
\hline
Diagnostic & $t/P_{\rm mb}=0.24$ & $0.30$ & $0.40$ & $0.64$ \\
Frame representation & moving & moving & moving & BH inertial \\
Domain size $(R_\odot^3)$ & $64^3$ & $256^3$ & $512^2\times256$ & $512^2\times256$ \\
$\Delta r_{\min}/R_\odot$ & $7.8\times10^{-3}$ & $1.6\times10^{-2}$ & $3.1\times10^{-2}$ & $6.3\times10^{-2}$ \\
$N_{\rm MB}$ & 2962 & 5573 & 4792 & 3833 \\
$(M_{\rm deb}-M_*)/M_*$ & $-5.6\times10^{-8}$ & $-8.3\times10^{-8}$ & $-3.4\times10^{-7}$ & $+1.1\times10^{-5}$ \\
$L_{{\rm I},z}/L_{{\rm I},z,0}$ & 1.0000 & 0.9995 & 0.9994 & 0.9983 \\
$f_{\rm bound}$ & 0.4928 & 0.5112 & 0.5119 & 0.5066 \\
\hline
\end{tabular}
\end{table*}

\subsection{Frame conversion}

When the calculation is converted to the BH inertial frame, the hydrodynamic variables are boosted using the current frame velocity,
\begin{align}
  \rho' &= \rho,\\
  (\rho{\bf v})' &= \rho({\bf v}+{\bf V}_{\rm fr}),\\
  {\cal E}_{\rm tot}' &= {\cal E}_{\rm tot}+\rho{\bf v}\cdot{\bf V}_{\rm fr}
       +\frac{1}{2}\rho |{\bf V}_{\rm fr}|^2 .
  \label{eq:galilean-boost}
\end{align}
The boost is skipped in floor-density cells so that the ambient medium does not acquire artificial kinetic energy. After the boost, the frame origin, velocity, and acceleration are reset to zero, and the BH is fixed in the grid coordinates. Subsequent domain enlargements use the same restart-remapping procedure in the BH inertial frame.

\subsection{Domain remapping of conserved variables}

We use the conserved hydrodynamical variables for remapping procedure,
\begin{equation}
  {\bf Q}_{\rm cons} =
  \left(\rho,\rho v_x,\rho v_y,\rho v_z,{\cal E}_{\rm tot},\ldots\right),
  \label{eq:remap-state-vector}
\end{equation}
where the ellipsis denotes any passive scalars and, when dual-energy evolution is enabled, the auxiliary internal energy is also involved. For each target cell $C$, the code constructs an approximate cell average by sampling the source AMR hierarchy at the physical coordinates of the target cell,
\begin{align}
  \overline{\bf Q}_{\rm cons,C}^{\,{\rm new}}
  &=
  \frac{1}{{\cal V}_C}
  \int_{{\cal V}_C}
  {\cal I}_{\rm AMR}
  \left[{\bf Q}_{\rm cons}^{\rm old}\right]({\bf r})\,d^3r
  \nonumber\\
  &\simeq
  \frac{1}{N_{\rm qd}}
  \sum_{\alpha=1}^{N_{\rm qd}}
  {\cal I}_{\rm AMR}
  \left[{\bf Q}_{\rm cons}^{\rm old}\right]({\bf r}_{C,\alpha}) .
  \label{eq:remap-cell-average}
\end{align}
Here ${\cal I}_{\rm AMR}$ denotes trilinear interpolation on the finest source MeshBlock that contains the sampling point, ${\cal V}_C$ is the target-cell volume, and ${\bf r}_{C,\alpha}$ are the $N_{\rm qd}$ sub-cell quadrature points. In the 3D remapping process, we adopt two midpoint samples per active coordinate direction with $N_{\rm qd}=2^3=8$.

The energy is carried across the remap as thermal energy rather than as total energy. We interpolate the density, momentum, and thermal energy density, using the dual-energy auxiliary variable where the source provides one, and then reconstruct the total energy of the target cell from these quantities. The kinetic energy of the interpolated momentum is always smaller than the kinetic energy before interpolation, and the difference is the kinetic energy of the velocity structure that the target cell cannot represent. Interpolating the total energy would convert this unresolved kinetic energy into heat, which is not negligible in a cold stream whose thermal energy is only $10^{-3}$ of its kinetic energy. Carrying the thermal energy instead discards this unresolved component and keeps the temperature and velocity within the range of the source values. Coincident cells are copied directly, so a remap onto the same grid is the identity operation. In a three-dimensional Orszag--Tang test remapped onto a grid twice as coarse, the total-energy form raises the internal energy by $0.96\%$, exactly the kinetic energy lost to coarsening, whereas the thermal-energy form changes it by only $1.8\times10^{-7}$.

Target points outside the old domain are set to the floor state, with density and pressure blended smoothly across a narrow band inside the numerical atmosphere, and cells inside the BH excision region are reset to the sink state.

The appropriate AMR hierarchy of the enlarged domain is not known in advance, so each remap projects the same source state twice. The first projection places the solution on the initial coarse grid only to build the AMR hierarchy during a short refinement-only settling phase, after which this temporary state is discarded. The source restart is then projected again onto the refined hierarchy, which avoids an initially under-resolved debris stream or BH neighborhood, and the normal AMR and output cadence resume.

\section{Dual-energy evolution}\label{sec:dual-energy}

The debris stream in a TDE is cold and highly supersonic over much of its evolution. In this regime, the internal energy is a small residual obtained by subtracting the kinetic energy from the conserved total energy. This subtraction is well-behaved in shocks and hot gas, where the thermal energy is dynamically important, but it can become inaccurate and even negative in nearly ballistic stream cells. A small truncation error in
\begin{equation}
  {\cal U}_{\rm cons}
  =
  {\cal E}_{\rm tot}-\frac{1}{2}\rho|{\bf v}|^2,
  \label{eq:dual-econs}
\end{equation}
can then produce an order-unity error in the temperature. We therefore follow the dual-energy strategy used in \texttt{Enzo} \citep{1995CoPhC..89..149B,2014ApJS..211...19B} and adapted in \texttt{Athena++} applications \citep{2022ApJ...941...73T,2024ApJ...963...26Z,2025ApJ...985...16T}. The conserved total-energy density ${\cal E}_{\rm tot}$ remains the finite-volume energy variable. In parallel, we evolve an auxiliary internal-energy density ${\cal U}_{\rm aux}$ as an additional hydrodynamical variable, and use it for EOS closure when the conservative subtraction in Eq.~(\ref{eq:dual-econs}) is unreliable.

\subsection{Effect in the production TDE simulation}
\label{sec:dual-energy-production}

\begin{figure*}
  \centering
  \includegraphics[width=.8\textwidth]{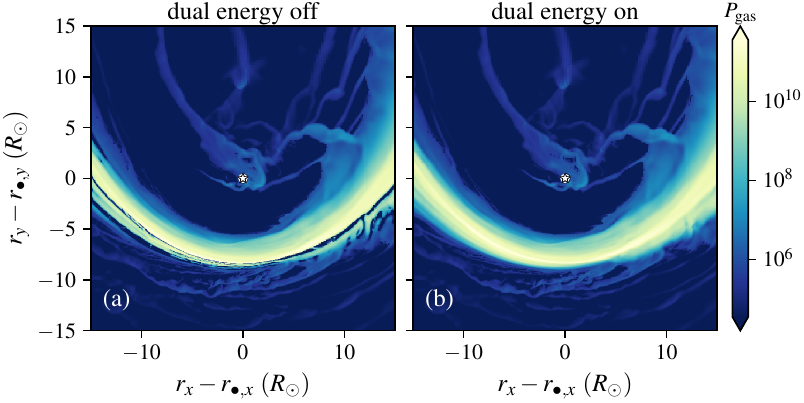}
  \caption{Dual-energy pressure validation in a production TDE state. Panels (a) and (b) show BH-centered midplane gas pressure in a
  $30R_\odot$-wide region around the BH for otherwise identical calculations with
  dual energy disabled and enabled. The shared colorbar gives $P_{\rm gas}$ in
  $\mathrm{dyn\,cm^{-2}}$.}
  \label{fig:dual-energy-pressure-validation}
\end{figure*}

Near pericenter, the kinetic energy of the debris stream can exceed the thermal energy by many orders of magnitude (Mach number $\mathcal{M}\gtrsim10^4$). Recovering pressure solely by subtracting kinetic energy from total energy then produces conserved-to-primitive (C2P) failures and activates the pressure floor, yielding the narrow low-pressure artifacts seen in Fig.~\ref{fig:dual-energy-pressure-validation}a. With the dual-energy approach, these floor-dominated structures disappear while the large-scale post-nozzle pressure morphology is retained.

\subsection{Dual-energy algorithm}

The auxiliary variable is initialized from the internal energy implied by the conserved state. During C2P conversion, we first compute ${\cal U}_{\rm cons}$ from Eq.~(\ref{eq:dual-econs}) and compare it with the local total energy. The internal-energy density used for EOS closure is
\begin{equation}
  {\cal U}_{\rm EOS} =
  \begin{cases}
    {\cal U}_{\rm cons}, &
    {\cal U}_{\rm cons}>0\ {\rm and}\ {\cal U}_{\rm cons}>\eta_1 {\cal E}_{\rm tot},\\
    {\cal U}_{\rm aux}, & \mathrm{otherwise}.
  \end{cases}
  \label{eq:dual-primitive-switch}
\end{equation}
The dimensionless parameter $\eta_1$ controls this pressure-recovery switch. The conservative branch (using ${\cal U}_{\rm cons}$) preserves the safely resolved shock-heated state. Otherwise, we use ${\cal U}_{\rm aux}$ to avoid noisy or negative pressures in cold, hypersonic cells. In the TDE runs presented here, we use $\eta_1=10^{-3}$.

The auxiliary internal energy is evolved in two operator-split stages. The first is a conservative advection stage. At each cell-surface, the auxiliary specific internal energy is upwinded and transported with the mass flux returned by the
Riemann solver,
\begin{equation}
  F_{{\cal U}_{\rm aux}} =
  F_\rho
  \left(\frac{{\cal U}_{\rm aux}}{\rho}\right)_{\rm upwind},
  \label{eq:dual-flux}
\end{equation}
where the upwind state is selected by the sign of the mass flux $F_\rho$. The same upwind state defines the normal cell-surface velocity in each direction,
\begin{equation}
    v_{f,d} = \frac{F_{\rho,d}}{\rho_{\rm upwind}},
    \qquad d\in\{x,y,z\},
    \label{eq:dual-face-velocity}
\end{equation}
which is stored for the second, non-conservative part of the auxiliary energy update.

After conservative advection, the second operator-split part accounts for compressional heating and expansion cooling. For the internal-energy density, this source is
\begin{equation}
  \frac{d {\cal U}_{\rm aux}}{dt}
  =
  -P\nabla\cdot{\bf v},
  \label{eq:dual-compression-continuous}
\end{equation}
where the time derivative denotes the operator-split source update after the advection step. At this stage, the cell volume is fixed, and the advective flux has already transported ${\cal U}_{\rm aux}$, thus, the compression of the advected density is not applied again. Only the remaining $-P\nabla\cdot{\bf v}$ pressure-work term is evaluated as a source. The discrete update uses the face-centered velocities from
Eq.~(\ref{eq:dual-face-velocity}) to form
\begin{align}
  (\nabla\cdot{\bf v})_{ijk}
  &=
  \frac{v_{f,x}|_{i+1/2}-v_{f,x}|_{i-1/2}}{\Delta x}
  \nonumber\\
  &\quad+
  \frac{v_{f,y}|_{j+1/2}-v_{f,y}|_{j-1/2}}{\Delta y}
  \nonumber\\
  &\quad+
  \frac{v_{f,z}|_{k+1/2}-v_{f,z}|_{k-1/2}}{\Delta z}.
  \label{eq:dual-divv}
\end{align}
For an ideal gas, $P=(\gamma-1){\cal U}_{\rm aux}$, and Eq.~(\ref{eq:dual-compression-continuous}) is integrated analytically over each stage with $\nabla\cdot{\bf v}$ held fixed, which keeps ${\cal U}_{\rm aux}$ positive. For a tabulated EOS, the pressure work is discretized directly,
\begin{equation}
  {\cal U}_{\rm aux}^{\,(\ell+1)} = {\cal U}_{\rm aux}^{\,*,(\ell)}
  - P\!\left(\rho^{(\ell)},{\cal U}_{\rm aux}^{\,*,(\ell)}/\rho^{(\ell)}\right)
  (\nabla\cdot{\bf v})^{(\ell)}\Delta t_\ell ,
  \label{eq:dual-tabulated-compression}
\end{equation}
where the superscript $(\ell)$ labels the RK stage, $\Delta t_\ell$ is its timestep, and ${\cal U}_{\rm aux}^{\,*,(\ell)}$ is the value after conservative advection in that stage. Thermal floors are applied before and after the EOS call so that the table is never evaluated outside its range.

The auxiliary variable is resynchronized with the conservative total energy wherever the conservative thermal energy is reliable. Following the same procedure as the dual-energy synchronization in \citet{1995CoPhC..89..149B}, we compare the recovered thermal energy with the total energy scale in the local neighborhood:
\begin{equation}
  \begin{aligned}
  {\cal U}_{\rm aux} &\leftarrow {\cal U}_{\rm cons}
  \quad \mathrm{if}\\
  {\cal U}_{\rm cons} &>0
  \quad \mathrm{and}\quad
  {\cal U}_{\rm cons}>
  \eta_2 \max_{\chi\in{\cal N}(i,j,k)} {\cal E}_{{\rm tot},\chi} .
  \end{aligned}
  \label{eq:dual-sync}
\end{equation}
Here ${\cal N}$ is the local neighborhood and for the production runs, we use $\eta_2=10^{-4}$. Unlike $\eta_1$, which selects the pressure closure, $\eta_2$ controls resynchronization so that reliable cells affected by shock-heating overwrite ${\cal U}_{\rm aux}$ while the cells in cold-stream remain protected.

\subsection{Validation in High-Mach Number Flow}

\begin{figure*}
  \centering
  \includegraphics[width=\textwidth]{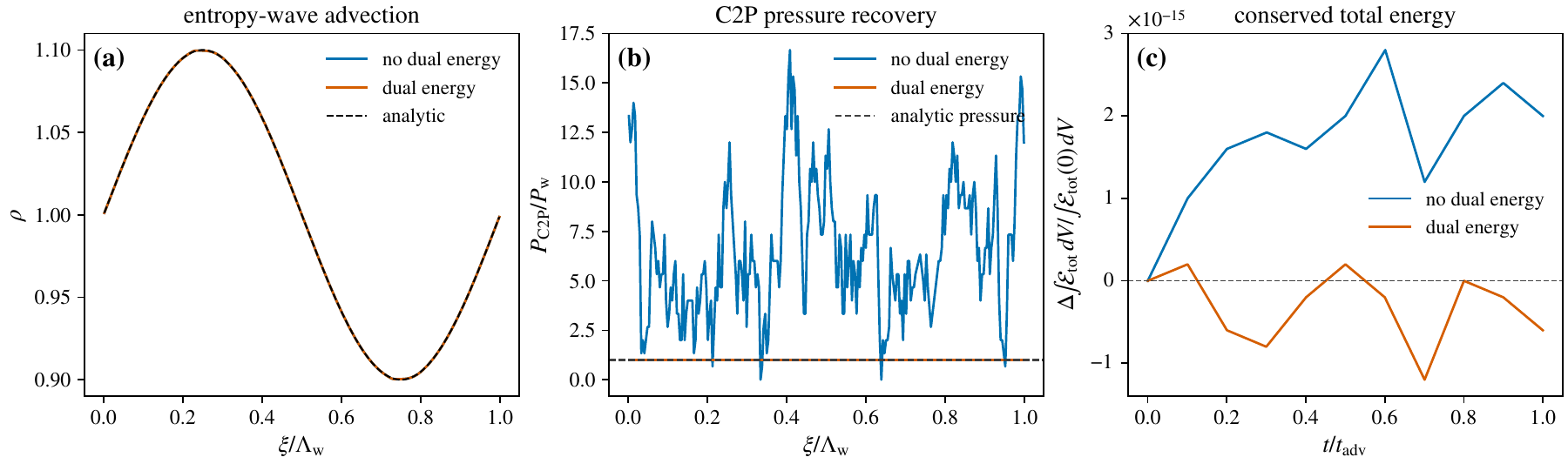}
  \caption{The C2P cancellation test in targeted high-Mach number flow after one advection period. (a) Numerical entropy-wave density compared to the exact solution. (b) Conservative subtraction alone yields order-unity pressure errors, whereas the dual-energy method recovers the exact $P=P_{\rm w}$ solution. (c) Volume-integrated total energy is conserved to roundoff precision in both calculations.}
  \label{fig:dual-energy-cancellation}
\end{figure*}

Fig.~\ref{fig:dual-energy-cancellation} presents a C2P test in a periodic entropy wave at high Mach number.

In this test, we use the periodic entropy wave
\begin{equation}
    \rho(\xi,t)=\bar\rho_{\rm w}
    \left[1+a_{\rm w}\sin\!\left(2\pi
    \frac{\xi-V_{\rm adv}t}{\Lambda_{\rm w}}\right)\right],
\end{equation}
where $\xi$ is the coordinate, $\bar\rho_{\rm w}$ the mean density, $a_{\rm w}$ the amplitude, and $\Lambda_{\rm w}$ the wavelength. The wave advects at a uniform speed $V_{\rm adv}=10^8$ (Mach number $\mathcal{M}=7.75\times10^7$) while pressure remains uniform at $P_{\rm w}$. For an ideal gas with $\gamma=5/3$, the maximum normalized pressure error $\delta_{P,\infty}=\max|P-P_{\rm w}|/P_{\rm w}$ after one period is $15.7$ without dual-energy treatment, indicating severe C2P failures. In contrast, the dual-energy method reduces $\delta_{P,\infty}$ to $4.0\times10^{-12}$ (machine precision) while maintaining total energy conservation to a relative precision of $6.0\times10^{-16}$.

\section{Localized Adaptive Time-stepping (LAT)}\label{sec:lat}

\subsection{LAT method}

The global timestep in an explicit finite-volume calculation is set by the most restrictive cell in the domain. In a TDE calculation, this restriction can become highly localized: a small number of cells near the black hole or in the nozzle shock demand a timestep far shorter than that required by the extended debris stream, even though the latter fills most of the MeshBlocks. We therefore implement localized adaptive time stepping (LAT) for the non-relativistic hydrodynamic update in \texttt{AthenaK}. Following the strategy used in \texttt{H-AMR} \citep{2022ApJS..263...26L}, MeshBlocks advance on power-of-two timestep bins, while synchronized boundary construction and time-integrated flux correction preserve finite-volume coupling between bins. In the matched production restart presented at the end of this section, this localization increases wall-clock throughput by a factor of $3.1$ while preserving the dense-stream morphology.

At a synchronized state, each MeshBlock estimates its local CFL limit from hydrodynamic and source terms, $\Delta t_b^{\rm CFL}$. Let
\begin{equation}
  \Delta t_0 = \min_b \Delta t_b^{\rm CFL}
  \label{eq:lat-base-dt}
\end{equation}
be the timestep that would be used by a globally synchronized calculation. Each block is assigned a timestep factor
\begin{equation}
  f_b =
  2^{\min\left[N_{\rm LAT},
  \left\lfloor \log_2
  \left(\frac{\Delta t_b^{\rm CFL}}{\Delta t_0}\right)
  \right\rfloor\right]},
  \qquad
  \Delta t_b = f_b \Delta t_0 ,
  \label{eq:lat-factor}
\end{equation}
where $N_{\rm LAT}$ is the configured number of LAT levels, which limits the factors to at most $2^{N_{\rm LAT}}$. After each synchronization, the factors are recomputed using the latest CFL estimates per-block, so that the timestep distribution tracks the evolving flow. Sparsely filled high‑factor bins are merged into the next smaller bin to prevent near‑empty bins from degrading parallel efficiency. 

The factors are then limited so that MeshBlocks sharing a face, edge, or corner differ by at most a factor of two, and $f_{\rm fine}\le f_{\rm coarse}\le2f_{\rm fine}$ across coarse--fine interfaces. This produces a gradual timestep staircase around rapidly evolving regions and keeps neighboring states time-compatible for AMR boundary filling and prolongation.

\begin{figure*}[t]
  \centering
  \includegraphics[width=0.94\textwidth]{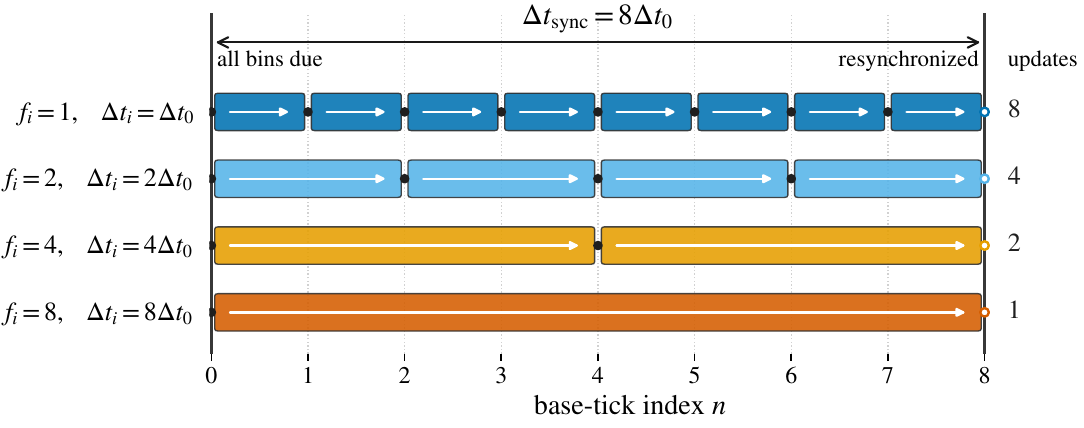}
  \caption{Schematic LAT schedule for a synchronization window with $f_{\max}=8$. Each colored arrow represents one MeshBlock update of duration $\Delta t_i=f_i\Delta t_0$; filled circles mark update launch times, and open circles mark the common synchronization time. Here $f_i$ is the timestep factor, not the AMR refinement level. The $f_i=1,2,4,$ and 8 blocks perform 8, 4, 2, and 1 updates within one window, respectively. For the four representative blocks shown, the ideal work count is therefore reduced from $4\times8=32$ updates with global synchronization to $8+4+2+1=15$ with LAT. Boundary-state reconstruction and communication overheads are not included in this count.}
  \label{fig:lat-schedule}
\end{figure*}

Let $f_{\max}\le2^{N_{\rm LAT}}$ be the largest timestep factor currently assigned. All MeshBlocks are synchronized at the end of a window of duration $\Delta t_{\rm sync}=f_{\max}\Delta t_0$. Within this window, a MeshBlock with factor $f_i$ advances with timestep $f_i\Delta t_0$ and therefore performs $f_{\max}/f_i$ updates. MeshBlocks scheduled at the same physical time are advanced as one active set, while boundary states for neighbors on different schedules are reconstructed from their stored time histories. Fig.~\ref{fig:lat-schedule} illustrates this power-of-two cadence for $f_{\max}=8$. A globally synchronized calculation performs $N_{\rm MB}f_{\max}$ MeshBlock updates per window, whereas LAT ideally performs $\sum_{i=1}^{N_{\rm MB}}f_{\max}/f_i$ updates.

Ghost states of neighbors on a different schedule are reconstructed at the required stage time from their stored endpoint states, using the dense output of the RK2 step where it is admissible and linear interpolation otherwise. At coarse--fine interfaces, the histories are restricted, interpolated in time, and then prolonged, so that all states entering the prolongation stencil refer to the same physical time.

\begin{figure*}
  \centering
  \includegraphics[width=.9\linewidth]{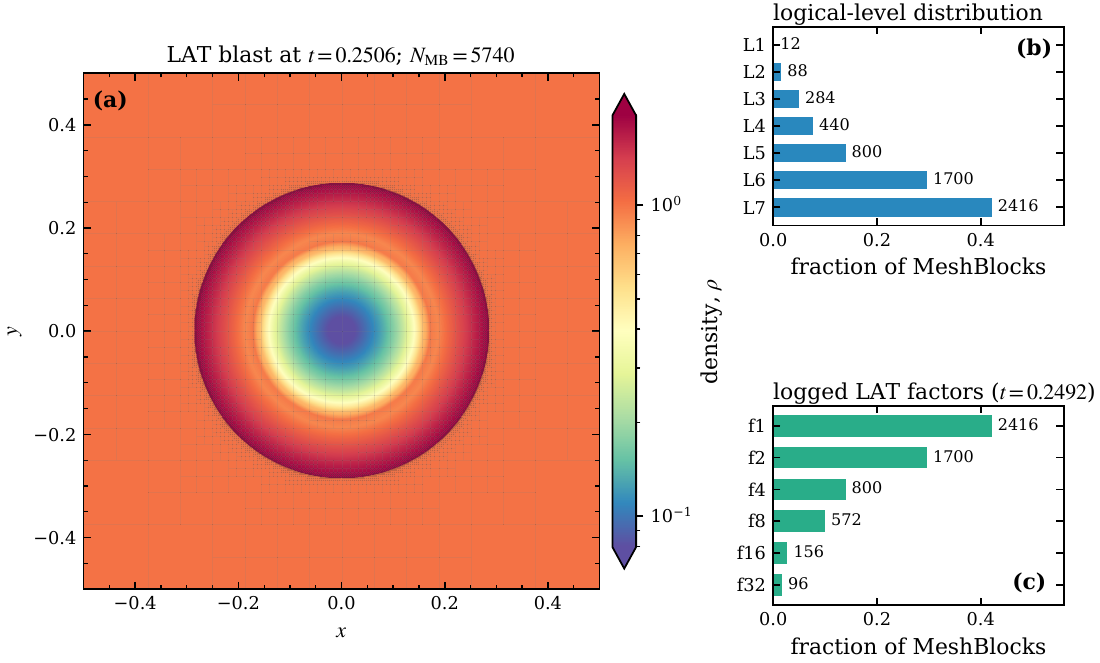}
  \caption{Mesh and timestep-bin structure of the LAT blast-wave validation run. Panel (a) shows the density and leaf MeshBlock boundaries at a representative output. Panels (b) and (c) show the logical-level distribution and timestep factor distribution, respectively. The finest and most frequently updated blocks follow the shock and contact structure, while the smooth ambient medium remains coarse and advances in larger timestep bins.}
  \label{fig:lat-blast-mesh}
\end{figure*}

\begin{figure}
  \centering
  \includegraphics[width=\columnwidth]{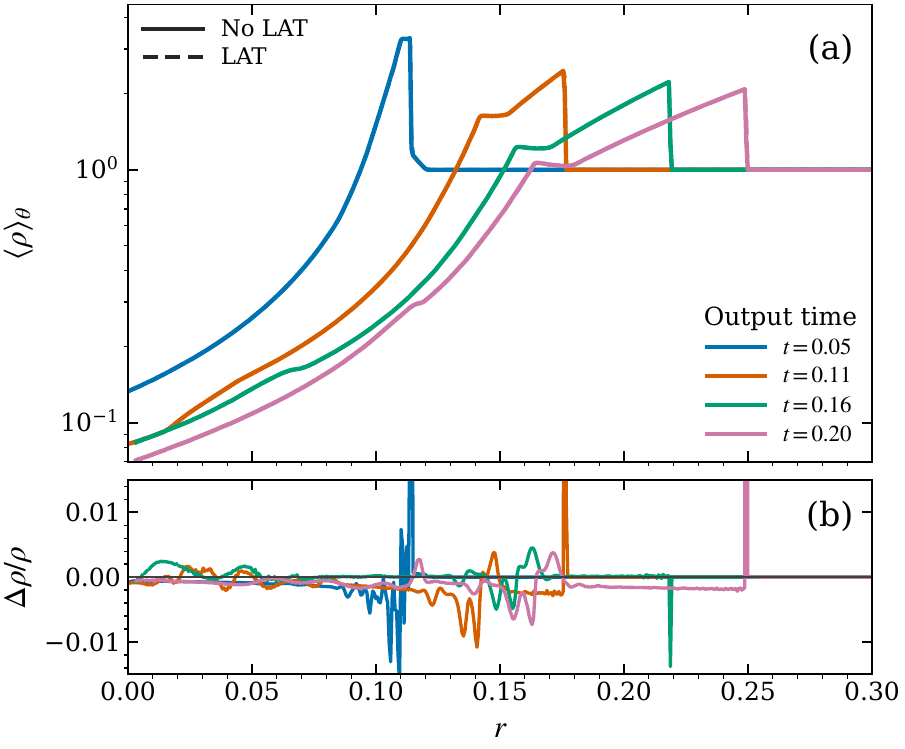}
  \caption{Hydrodynamic LAT validation using a two-dimensional blast wave propagation in AMR. Panel (a) compares azimuthally averaged density profiles from the globally synchronized (no-LAT) calculation (solid lines) and LAT calculation (dashed lines) at four output times. Panel (b) shows the relative density difference $\Delta\rho/\rho=(\rho_{\rm LAT}-\rho_{\rm noLAT})/\rho_{\rm noLAT}$.}
  \label{fig:lat-blast-validation}
\end{figure}

Conservation across mixed-timestep interfaces is enforced by delayed recalculation of numerical flux at the cell-surface, following the refluxing of block-structured AMR \citep{1989JCoPh..82...64B} and the strategy used in \texttt{H-AMR} \citep{2022ApJS..263...26L}. A fast MeshBlock with a larger time step may evaluate a shared cell-surface several times before its slower neighbor MeshBlock with a smaller time step reaches the same completion time. The fast MeshBlock therefore accumulates the time-integrated numerical flux from the slow MeshBlock
\begin{equation}
  {\cal I}_{\rm face} =
  \sum_\nu {\bf F}^{(\nu)} {\cal A}_{\rm f}\,\Delta t_\nu ,
  \label{eq:lat-flux-impulse}
\end{equation}
where ${\cal A}_{\rm f}$ is the cell-surface area, $\nu$ labels completed sub-steps, and ${\bf F}^{(\nu)}$ is the finite-volume face flux. At each compatible completion phase, the accumulated numerical flux replaces the inconsistent neighbor contribution and is deposited with the appropriate sign and volume normalization. The correction is applied at both AMR coarse--fine interfaces and mixed-factor same-level cell-surfaces, preserving conservation across the asynchronous update.

Load balancing gives each GPU a comparable mix of frequently and rarely updated MeshBlocks, so that the work per fine sub-step is equalized, and it is recomputed at synchronization points after the AMR hierarchy changes.

LAT is optional and reduces to the standard \texttt{AthenaK} update when all factors are unity. Its intended regime is a flow with a broad but spatially localized timestep spread, such as a long TDE debris stream whose smallest timesteps occupy only a small part of the MeshBlock.

\subsection{Coupling with gravity}
Gravity complicates LAT because the source term is not a purely local hydrodynamic flux. The total potential combines an analytical BH potential and a numerical self-gravity potential, as defined in Eq.~(\ref{eq:potential-decomposition}). Although the update from local gravity source term in Eqs.~(\ref{eq:lat-grav-momentum})--(\ref{eq:lat-grav-energy}) remains unchanged, LAT alters the temporal coupling. Active MeshBlocks must use a potential synchronized with their sub-step time, and the correction from recalculated numerical flux must incorporate the gravitational work. 

Sub-cycling with self-gravity is well established when the timestep is set per refinement level rather than per MeshBlock, as in \texttt{Enzo} \citep{2014ApJS..211...19B} and \texttt{CASTRO} \citep{2010ApJ...715.1221A}, where the Poisson equation is solved level by level. Under LAT, the timestep factor is assigned per MeshBlock, so blocks on the same refinement level can be at different physical times within a window. There is therefore no level at which a globally consistent potential can be constructed, which is why \citet{2023ApJS..266....7T} adopt uniform time stepping. We retain the globally consistent solve but restrict it to the synchronized states between windows and correct the gravitational work at the timestep interfaces.

The source term is updated only for active MeshBlocks during the current LAT sub-step. The analytical BH potential is evaluated directly from the instantaneous BH position and mass at the sub-step time. In contrast, the numerical self-gravity potential $\Phi_{\rm sg}({\bf r},t_{\rm sg})$ remains frozen between successive multigrid solves. To control this approximation, the length of a LAT window is limited to the refresh interval $\Delta t_{\rm sg}$. If a window would cross the next scheduled refresh, the solve is triggered at its start. The potential is therefore never older than $\Delta t_{\rm sg}$, and the Poisson equation is solved only at globally synchronized states where every MeshBlock is at the same physical time.

At a mixed-timestep interface, the accumulated numerical flux from the neighbor cells with slower timestep from Eq.~(\ref{eq:lat-flux-impulse}) determines the mass, momentum, and hydrodynamic energy that must be recalculated. Gravity adds an additional energy contribution because the recalculated mass flux has crossed a potential difference. Let
\begin{equation}
  \delta {\cal I}_{\rho,f}
  = {\cal I}_{\rho,f}^{\rm local}
  - {\cal I}_{\rho,f}^{\rm accepted}
  \label{eq:lat-mass-impulse-mismatch}
\end{equation}
be the area- and time-integrated mass-flux mismatch retained on cell-surface $f$
after the accepted fine-side or mixed-factor flux has been received. The
gravitational work correction is applied before the recalculation of ordinary conserved-variable numerical flux. For receiving cell $i$,
\begin{equation}
  \delta {\cal E}_{{\rm tot},i}^{\rm grav,ref}
  =
  \frac{\sigma_{i,f}}{2{\cal V}_i}\,
  \delta {\cal I}_{\rho,f}
  \left[\Phi_{n(f)}-\Phi_i\right],
  \label{eq:lat-grav-reflux}
\end{equation}
where $n(f)$ is the cell across surface, ${\cal V}_i$ is the receiving-cell volume, and $\sigma_{i,f}=\pm1$ is its outward-surface orientation.
The factor $1/2$ is the same symmetric potential-difference factor used in
Eq.~(\ref{eq:lat-grav-energy}). Eq.~(\ref{eq:lat-grav-reflux}) is used without change at the same level surfaces of coarse-fine and mixed-factor. The potential includes both the analytical BH and frozen numerical self-gravity contributions evaluated at synchronization. 

\subsection{Validation of LAT}

We verify the hydrodynamic LAT implementation with two test problems. The first is a two-dimensional spherical blast wave on an AMR hierarchy with 8 refinement levels, evolved both with the standard globally synchronized update and with LAT enabled. This problem contains strong shocks and moving AMR boundaries without gravity, which isolates the LAT boundary interpolation, active-MeshBlock scheduling, and correction of delayed numerical flux described above. Both calculations use the same ideal-gas equation of state, Riemann solver, reconstruction, AMR criterion, hierarchy, and refinement interval. The LAT run differs only in the local timestep schedule and the associated mixed-timestep synchronization.

Fig.~\ref{fig:lat-blast-mesh} and \ref{fig:lat-blast-validation} show how the spatial refinement and time-step distribution respond to the evolving blast wave. AMR tracks the shock and contact discontinuity. The timestep distribution does not simply mirror the AMR level, because local CFL conditions and neighbor limits also dictate the update frequency. The azimuthally averaged profiles remain closely matched between LAT and globally synchronized runs, with residuals localized around moving features. Although LAT reduces the update work by $41.3\%$, the rapidly moving refinement pattern and small active sets expose communication overhead. Consequently, the wall time increases by $44\%$, making this calculation a correctness and synchronization stress test rather than a performance benchmark.

We next test the coupling of LAT to self-gravity using the EOS-balanced star. Both runs share identical setups except for the hydrodynamic schedule, refreshing the potential every $\Delta t_{\rm sg}=0.03$$\,t_0$. Because the star is nearly hydrostatic, secular drift provides a sensitive test of the gravity work correction in Eq.~(\ref{eq:lat-grav-reflux}). Over an evolution of $16.35\,t_{\rm dyn}$, the final density profile errors are $4.01\times10^{-3}$ without LAT and $6.11\times10^{-3}$ with LAT. The stellar mass drift remains nearly identical at $\sim3.4\times10^{-5}$, and the normalized kinetic energy stays below $5.7\times10^{-6}$ in both cases. This close agreement demonstrates that the LAT gravity coupling remains stable.

\begin{figure*}
  \centering
  \includegraphics[width=\textwidth]{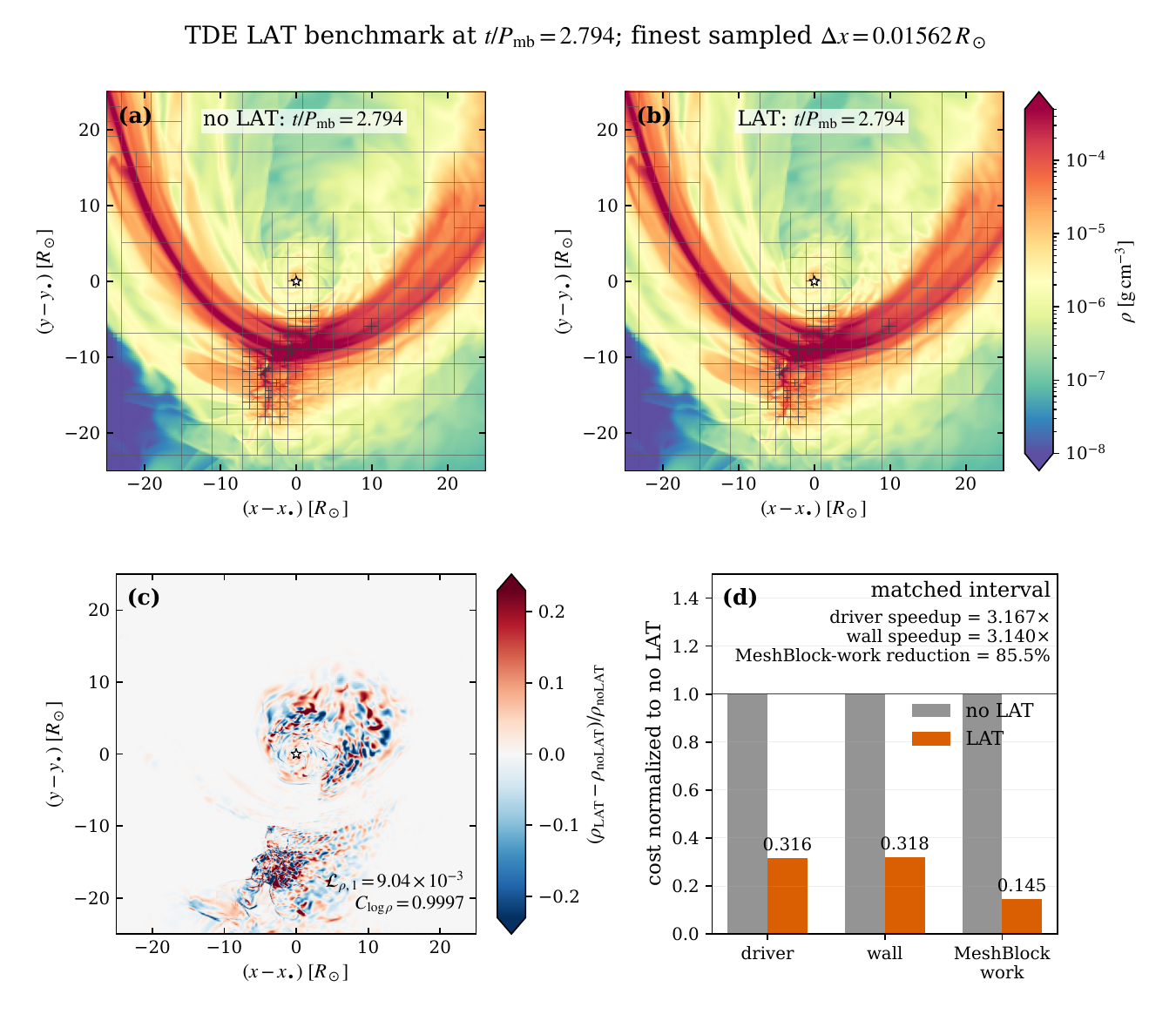}
  \caption{Realistic TDE simulation restart benchmark at matched $t/P_{\rm mb}=2.794$. Panels (a) and (b) show BH-centered density slices from globally synchronized and LAT schedules, panel (c) shows their relative density difference, and panel (d) gives the measured MeshBlock-work and wall-speed ratios. Gray outlines in the density panels show the AMR MeshBlock decomposition. Both runs use the same EOS, self-gravity, AMR hierarchy, and
  target epoch.}
  \label{fig:current-tde-lat-benchmark}
\end{figure*}

\subsection{Production TDE benchmark}
\label{sec:lat-benchmark}

To evaluate LAT under production conditions, we restart the \texttt{HR} calculation twice from a common checkpoint at $t/P_{\rm mb}=2.737$, once with the globally synchronized update and once with LAT with seven levels and $f_{\max}=128$, keeping every other setting identical. Both restarts are evolved for five code time units to $t/P_{\rm mb}=2.794$, which spans about $5\times10^4$ fine sub-steps and hundreds of LAT synchronization windows.

Fig.~\ref{fig:current-tde-lat-benchmark} compares the two solutions at their final simulation time. To quantify the agreement, both solutions are sampled on the same uniform sampling grid covering the region shown in Fig.~\ref{fig:current-tde-lat-benchmark}.
We define the pointwise relative density difference as
\begin{equation}
  \delta_{\rho,i}
  =
  \frac{\rho_{{\rm LAT},i}-\rho_{{\rm sync},i}}
       {\rho_{{\rm sync},i}},
\end{equation}
and the corresponding norms as
\begin{equation}
\begin{aligned}
  \mathcal L_{\rho,1}
  &= \frac{1}{N}\sum_{i=1}^{N}|\delta_{\rho,i}|,\\
  \mathcal L_{\rho,2}
  &= \left(\frac{1}{N}\sum_{i=1}^{N}
     |\delta_{\rho,i}|^2\right)^{1/2},\\
  \mathcal L_{\rho,\infty}
  &= \max_i|\delta_{\rho,i}|.
\end{aligned}
\label{eq:lat-density-error-norms}
\end{equation}
The normalized density differences are $\mathcal L_{\rho,1}=9.04\times10^{-3}$, $\mathcal L_{\rho,2}=2.94\times10^{-2}$, and $\mathcal L_{\rho,\infty}=1.64\times10^{-1}$. We further define the log-density correlation as the Pearson coefficient of $q\equiv\log_{10}\rho$ over the same $N$ sample points, $C_{\log\rho}=\sum_i(q_{{\rm LAT},i}-\bar q_{\rm LAT})(q_{{\rm sync},i}-\bar q_{\rm sync})/[\sum_i(q_{{\rm LAT},i}-\bar q_{\rm LAT})^2\sum_i(q_{{\rm sync},i}-\bar q_{\rm sync})^2]^{1/2}$, and the dense-region overlap as the Jaccard index $J_{\rm dense}=|D_{\rm LAT}\cap D_{\rm sync}|/|D_{\rm LAT}\cup D_{\rm sync}|$ of the sets of sample points $D$ with $\rho>10^2\rho_{\rm floor}$. The log-density correlation is $C_{\log\rho}=0.99975$, and the dense-region overlap metric is $J_{\rm dense}=1$. The largest pointwise residuals that set $\mathcal L_{\rho,\infty}$ arise in the tenuous, turbulent gas outside the main debris stream. Small perturbations from the differing time-step schedules are amplified there by chaotic nonlinear evolution, thus the detailed low-density morphology is decorrelated. In contrast, the dense, ordered stream remains closely matched, with no systematic LAT-induced displacement or broadening.

The benchmark ran on one node with ten 16 GB NVIDIA V100 GPUs, one MPI rank per GPU. LAT reduced the number of MeshBlock updates by 85\% and increased the wall-clock throughput by a factor of 3.1. The speedup is smaller than the reduction in update work because communication, synchronization, and task-launch overheads do not decrease in proportion to the number of active MeshBlocks, and the slightly different AMR update frequencies change the final number of MeshBlocks by less than 5\%. At the final time, the integrated mass agrees to the reported precision, and the volume-integrated momentum and total energy differ by $4\times10^{-6}$ and $1\times10^{-5}$, respectively.

\section{Integrated TDE validation}
\label{sec:integrated-tde-validation}

\subsection{Self-gravity cadence during the disruption}
\label{sec:selfgrav-cadence}

\begin{figure*}
  \centering
  \includegraphics[width=\textwidth]{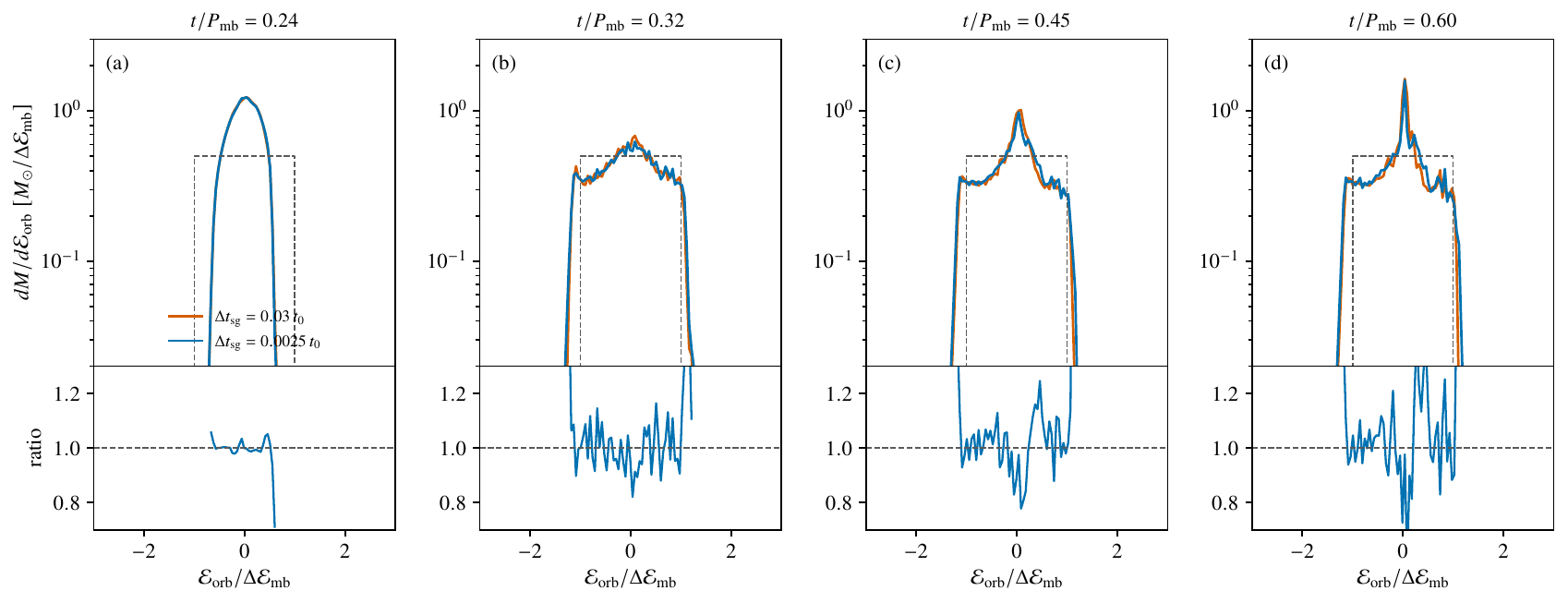}
  \caption{Debris energy distribution computed with self-gravity cadences $\Delta t_{\rm sg}=0.03\,t_0$ (production) and $0.0025\,t_0$, at pericenter (a) and at three later epochs (b)--(d). The dashed line is the top-hat reference, and the lower sub-panels show the ratio of the two distributions.}
  \label{fig:selfgrav-cadence}
\end{figure*}

The debris energy distribution is established while the star is torn apart, so the self-gravity cadence matters most around the first pericenter passage. We repeated the production sequence from the initial star through pericenter ($t\simeq0.24\,P_{\rm mb}$) to $t=0.6\,P_{\rm mb}$, including the first three domain remaps and the conversion to the BH inertial frame, once with the production cadence $\Delta t_{\rm sg}=0.03\,t_0$ and once with $\Delta t_{\rm sg}=0.0025\,t_0$, 12 times smaller. All other parameters are identical. Fig.~\ref{fig:selfgrav-cadence} compares the resulting debris energy distributions. At every epoch, the bound fractions differ by less than $0.004$ and the widths of the energy interval containing $90\%$ of the mass by about $2\%$. The cumulative mass distributions $M(<{\cal E}_{\rm orb})/M$ differ by at most $1.6\%$ on the bound side, which sets the fallback rate, and by at most $4\%$ overall. The largest difference lies at ${\cal E}_{\rm orb}\simeq+0.2\,\Delta{\cal E}_{\rm mb}$, next to the dense peak discussed in Sec.~\ref{sec:debris-energy}. The median mass-weighted vertical thickness of the debris agrees to $4\%$. The debris energy distribution is therefore insensitive to the self-gravity cadence over this range.

\subsection{Debris energy distribution and ballistic fallback}
\label{sec:debris-energy}

\begin{figure*}
  \centering
  \includegraphics[width=\textwidth]{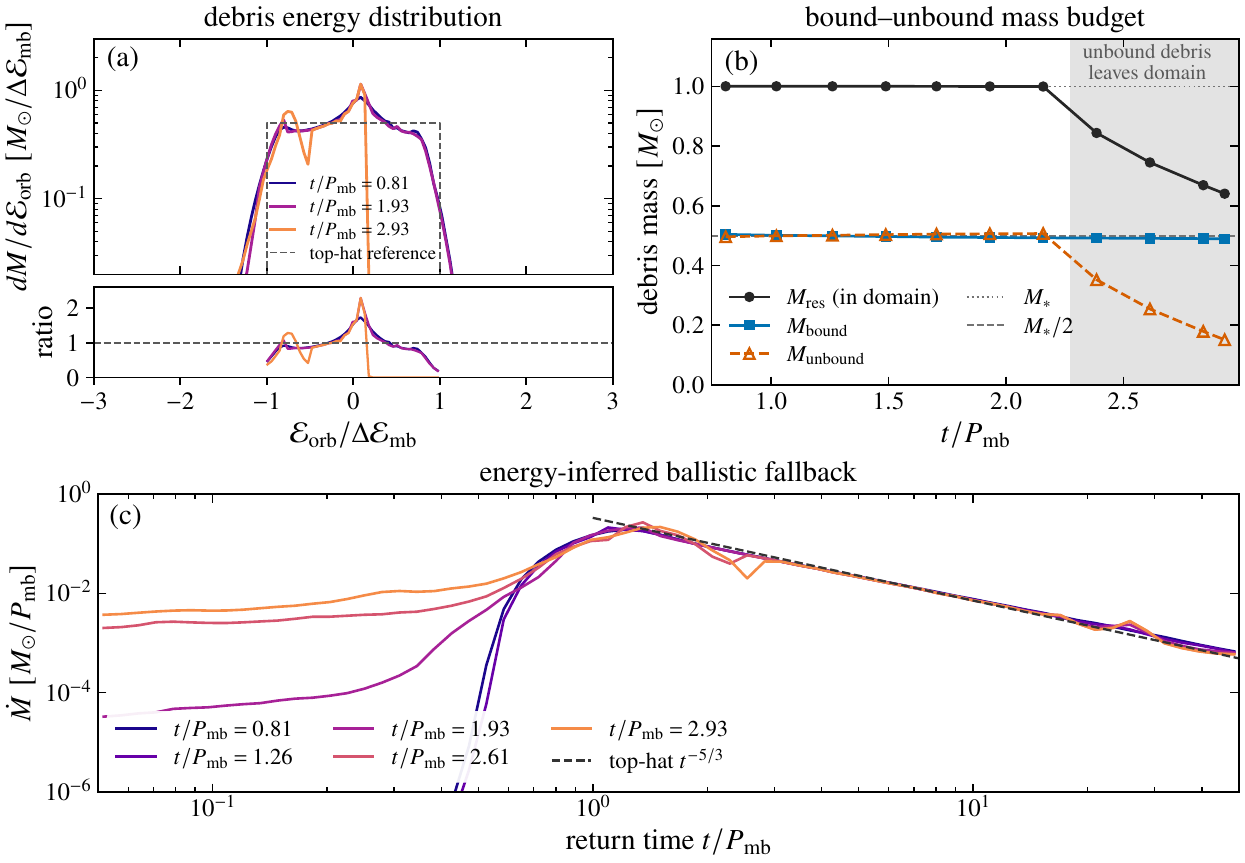}
  \caption{Comparison of the \texttt{HR} run with standard TDE energy distribution and fallback benchmarks. (a) Signed debris distribution $dM/d{\cal E}_{\rm orb}$ at three representative epochs, colored from dark to light, on a linear energy axis. The dashed line marks the uniform reference with $dM/d{\cal E}_{\rm orb}=M_*/(2\Delta{\cal E}_{\rm mb})$ for $|{\cal E}_{\rm orb}|\le\Delta{\cal E}_{\rm mb}$ and zero outside. The lower sub-panel shows the ratio to this top-hat value. At the last epoch, the unbound debris with ${\cal E}_{\rm orb}\gtrsim0.2\,\Delta{\cal E}_{\rm mb}$ has left the domain. (b) Resolved mass $M_{\rm res}$ ($\equiv M_{\rm deb}$ of Table~\ref{tab:conservation-audits}) and its bound (${\cal E}_{\rm orb}<0$) and unbound components. Dotted and dashed horizontal lines mark $M_*$ and $M_*/2$. The shaded band marks when unbound material begins crossing the outer boundary, causing $M_{\rm res}$ and $M_{\rm unbound}$ to drop together while $M_{\rm bound}$ stays flat. (c) Ballistic fallback rates inferred from the bound energy distributions, with the normalized $t^{-5/3}$ prediction shown dashed line. These rates are derived from the energy distribution, not direct measurements of the BH accretion rate.}
  \label{fig:integrated-tde-validation}
\end{figure*}

We now compare the production run \texttt{HR} with the standard debris-energy distribution, bound and unbound mass partition, and ballistic fallback rate shown in Fig.~\ref{fig:integrated-tde-validation} \citep{2021ARA&A..59...21G,2021SSRv..217...40R,2020ApJ...904...98R,2020ApJ...904...99R}.

Here the subscript ${\rm I}$ denotes components expressed in the BH inertial frame, while $\bullet$ labels the BH. For each output, the specific orbital energy relative to the BH is given by
\begin{equation}
  {\cal E}_{\rm orb} = \frac{1}{2}|{\bf v}_{\rm I}-{\bf v}_{\bullet,{\rm I}}|^2
  -\frac{G M_\bullet}{\sqrt{|{\bf R}-{\bf R}_\bullet|^2+r_{\rm soft}^2}},
  \label{eq:tde-orbital-energy}
\end{equation}
and its mass distribution is $dM/d{\cal E}_{\rm orb}$. In the frozen-in impulse picture, the BH tidal field produces an approximately antisymmetric energy spread about the center-of-mass orbital energy. 
For an initially parabolic orbit, the characteristic energy scale is
\begin{equation}
  \Delta{\cal E}_{\rm mb}=\frac{G M_\bullet R_*}{r_{\rm t}^2}.
  \label{eq:tde-most-bound-energy}
\end{equation}
It is the spread of the BH potential across the star at the tidal radius \citep{2021SSRv..217...40R,2020ApJ...904...98R}.
We use $\Delta{\cal E}_{\rm mb}$ as a normalization scale rather than assuming an instantaneous freeze-in of the debris energy distribution. 
A commonly used idealized reference is the top-hat distribution,
\begin{equation}
  \left(\frac{dM}{d{\cal E}_{\rm orb}}\right)_{\rm top\mbox{-}hat}
  =
  \begin{cases}
    M_*/(2\Delta{\cal E}_{\rm mb}), &
    |{\cal E}_{\rm orb}|\leq\Delta{\cal E}_{\rm mb},\\
    0, & |{\cal E}_{\rm orb}|>\Delta{\cal E}_{\rm mb}.
  \end{cases}
  \label{eq:tde-top-hat-energy}
\end{equation}
The calculations of hydrodynamical disruption generally produce a broadly flat interior distribution, although deviations from this idealized form can arise from self-gravity and subsequent hydrodynamic evolution of the debris
\citep{2020ApJ...904...98R,2020ApJ...904...99R,2021ApJ...923..184N}.

Panel~(a) shows that our debris distribution has the expected width and a rapid rolloff near $|{\cal E}_{\rm orb}|\sim\Delta{\cal E}_{\rm mb}$, but it is not flat. Relative to the top-hat value, it is $1.1$--$1.7$ for $|{\cal E}_{\rm orb}|\lesssim0.3\,\Delta{\cal E}_{\rm mb}$ and falls to $0.5$ at $\Delta{\cal E}_{\rm mb}$. At $t/P_{\rm mb}=1.70$, gas denser than $2\times10^{-3}\,{\rm g\,cm^{-3}}$ holds $21\%$ of the mass and has a mean orbital energy of $+0.07\,\Delta{\cal E}_{\rm mb}$. Such a dense central concentration is consistent with $\beta=1$ lying only slightly above the critical penetration factor $\beta\simeq0.9$ for the full disruption of a $\gamma=5/3$ polytrope \citep{2013ApJ...767...25G}, which our stellar model closely resembles (Sec.~\ref{sec:eos-star}). The weak evolution among the three plotted curves indicates that the debris energy distribution has largely frozen in by these epochs. However, as shocks and dissipation develop, the most tightly bound material begins to form substructures, deviating from purely ballistic motion.

For a complete disruption on an initially parabolic orbit, a symmetric energy spread places roughly half of the debris on bound orbits (${\cal E}_{\rm orb}<0$) and half on unbound orbits (${\cal E}_{\rm orb}>0$) \citep{2021ARA&A..59...21G,2021SSRv..217...40R,2020ApJ...904...99R}. Panel~(b) shows that the measured bound fraction stays between $0.495$ and $0.504$, in close agreement with the standard theory. 

The Keplerian energy--period relation provides the bound debris distribution to a ballistic mass-return rate
\citep{2021ARA&A..59...21G,2021SSRv..217...40R,
2020ApJ...904...98R,2020ApJ...905..141L}:
\begin{equation}
\begin{aligned}
  t_{\rm ret}({\cal E}_{\rm orb})
  &=2\pi G M_\bullet(2|{\cal E}_{\rm orb}|)^{-3/2},\\
  \dot M_{\rm fb}(t)
  &=\left.\frac{dM}{d|{\cal E}_{\rm orb}|}\right|_{{\cal E}_{\rm orb}(t)}
    \left|\frac{d|{\cal E}_{\rm orb}|}{dt}\right|.
\end{aligned}
\label{eq:tde-fallback-map}
\end{equation}
For Eq.~(\ref{eq:tde-top-hat-energy}), this becomes
\begin{equation}
  \dot M_{\rm fb}(t)=\frac{M_*}{3P_{\rm mb}}
  \left(\frac{t}{P_{\rm mb}}\right)^{-5/3}.
  \label{eq:tde-fallback-five-thirds}
\end{equation}
Using the mass distributions from our simulation, we construct the ballistic fallback rate shown in Panel~(c). The curves display a smoother turn-on and a lower, broader peak near the return time of the most-bound debris, reflecting the detailed structure visible in Panel~(a). At later times, the results from different output epochs converge and approach the normalized $t^{-5/3}$ reference, confirming that the bound side of the distribution near ${\cal E}_{\rm orb}\simeq0$ is close to the top-hat level and temporally stable. We stress that this is an energy-inferred ballistic fallback rate, not a direct measurement of the sink flux or BH accretion rate, and that the finite duration of the simulation does not constitute an asymptotic late-time fallback calculation.

\subsection{Pericenter nozzle shock: physical closure and resolution}
\label{sec:nozzle}

\begin{figure*}
  \centering
  \includegraphics[width=\textwidth]{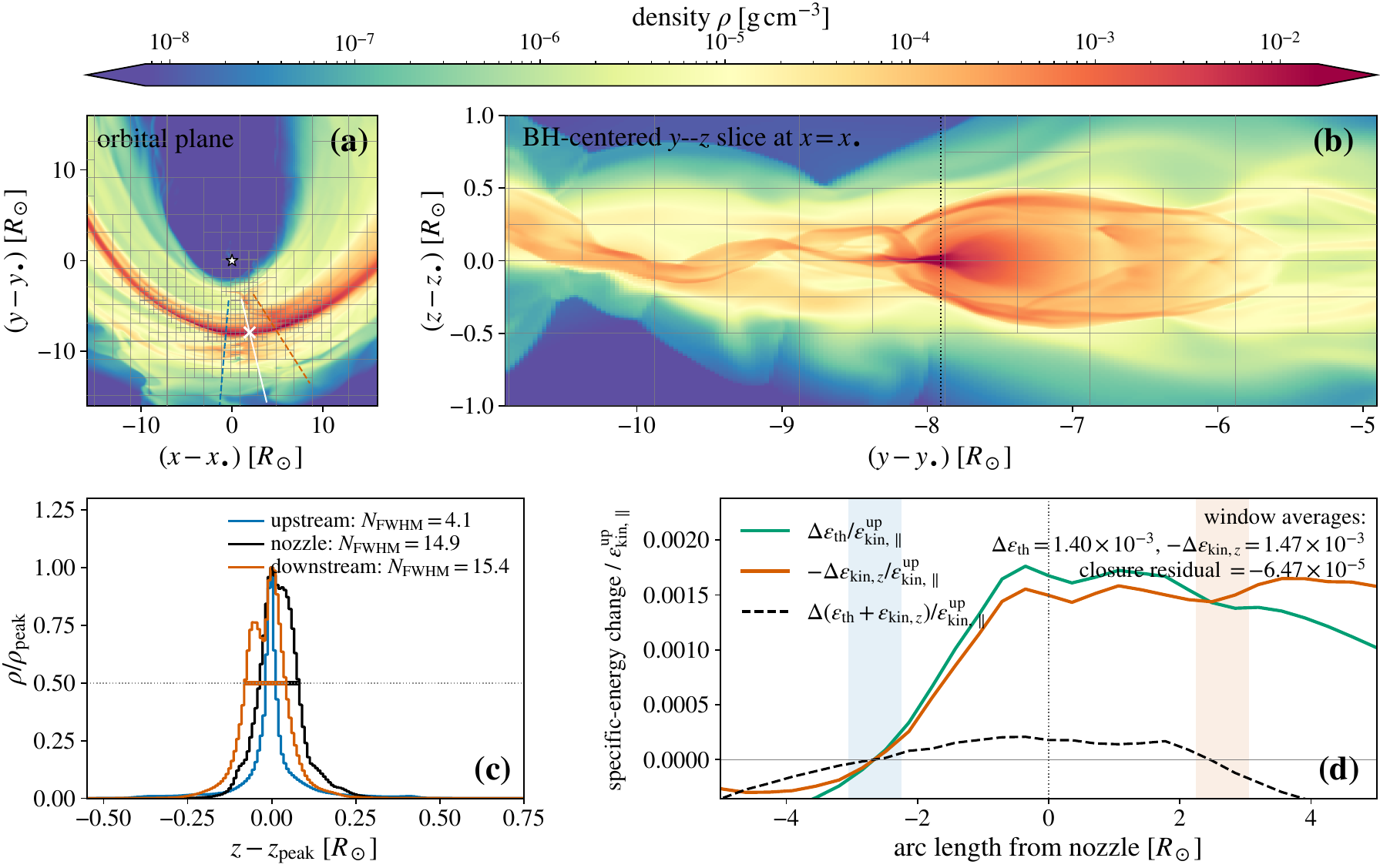}
\caption{Local nozzle shock resolution and vertical energy diagnostic at $t/P_{\rm mb}=1.70$ from the \texttt{HR} production calculation. Panel (a) shows a magnified orbital plane density view centered on the BH with overlaid AMR MeshBlock layouts. The white line marks the nozzle azimuth $\phi_{\rm noz}$. The blue and orange dashed lines mark the upstream and downstream sampling planes, respectively. Panel (b) presents a vertical density slice in the $y-z$ plane at $x=x_\bullet$, which passes $1.9\,R_\odot$ from the nozzle. The vertical dotted line marks the $y$ coordinate of the nozzle. Panel (c) gives the measured vertical density profiles and their half-maximum widths. These widths correspond to 4.1, 14.9, and 15.4 local AMR cells upstream, at the nozzle, and downstream, respectively. Panel (d) compares the specific thermal energy increase with the loss of vertical specific kinetic energy, with a window-averaged ratio $\Delta\varepsilon_{\rm th}/(-\Delta\varepsilon_{{\rm kin},z})=0.96$.}  \label{fig:nozzle-energy-resolution}
\end{figure*}

\citet{2026arXiv260520327N} identified a specific failure mode in which the longitudinal velocity transits from divergence to convergence near the pericenter. This occurs even though the smooth ballistic flow in the orbital plane lacks a physical caustic. An inadequately resolved shock-capturing calculation can erroneously convert this convergence into anomalous heating and a spurious fan. We evaluate this numerical artifact separately from the physical vertical compression that produces the nozzle shock. For this test, we use the high-resolution run \texttt{HR} at $t/P_{\rm mb}=1.70$ when the nozzle shock first forms and reaches maximum strength. Fig.~\ref{fig:nozzle-energy-resolution} shows the analysis.

As in the previous analysis, all measurements here are evaluated in the BH inertial frame. We define the relative velocity as $\delta{\bf v}\equiv{\bf v}_{\rm I}-{\bf v}_{\bullet,{\rm I}}$. The corresponding orbital plane and vertical specific kinetic energies are $\varepsilon_{{\rm kin},\parallel}=\tfrac{1}{2}(\delta v_x^2+\delta v_y^2)$ and $\varepsilon_{{\rm kin},z}=\tfrac{1}{2}\delta v_z^2$ respectively. The specific thermal energy $\varepsilon_{\rm th}={\cal U}_{\rm EOS}/\rho$ is obtained directly from the tabulated EOS closure. Mass-weighted angular profiles select the stream in a BH-centered annulus that brackets the nozzle, where $\phi$ is the orbital azimuth in the Cartesian $xy$ plane.

To analyze the stream around the pericenter nozzle, we first select cells where the density exceeds $10^2\rho_{\rm floor}$ and divide the domain into 144 equal azimuthal bins with $\Delta\phi=2.5^\circ$. Bins containing less than 5\% of the peak angular bin mass are excluded. In each remaining bin, the vertical stream thickness is evaluated as the mass-weighted root-mean-square $H_z=[\langle(z-z_\bullet)^2\rangle_M-\langle z-z_\bullet\rangle_M^2]^{1/2}$. The nozzle center $\phi_{\rm noz}$ is identified as the bin with the minimum $H_z$, corresponding to the strongest vertical compression, and its radius $r_{\rm noz}$ is defined by the mass-weighted cylindrical radius $\langle R\rangle_M$ in that bin. This procedure yields $\phi_{\rm noz}=-76.25^\circ$ and $r_{\rm noz}=8.14\,R_\odot$. To evaluate energy budgets across the nozzle shock, we select dense stream cells within $4.4\le R/R_\odot\le18$ and $|z-z_\bullet|\le6\,R_\odot$, and define symmetric upstream and downstream windows covering $0.20\le|\phi-\phi_{\rm noz}|\le0.45$~rad. Here, $\langle\cdot\rangle_M$ denotes a mass weighted cell average, $\Delta\varepsilon_X$ represents the difference between downstream and upstream values, and $\varepsilon_{{\rm kin},\parallel}^{\rm up}$ is the upstream parallel kinetic energy.

Between the upstream and downstream windows, the specific thermal energy rises by $1.40\times10^{-3}\,\varepsilon_{{\rm kin},\parallel}^{\rm up}$, while the vertical kinetic energy falls by $1.47\times10^{-3}\,\varepsilon_{{\rm kin},\parallel}^{\rm up}$ with a ratio of $0.96$. Therefore the heating is consistent with the loss of vertical kinetic energy at the nozzle.

Total energy conservation alone does not exclude spurious shock heating reported by \citet{2026arXiv260520327N}. We additionally examine the entropy generation following their approach. During the first pericenter passage, the star remains intact and should evolve adiabatically. Over this interval, the measured changes in the mass-weighted specific entropy are only $-0.02$ and $+0.04\,k_{\rm B}$ per baryon despite the Mach number exceeding $\mathcal{M}>10^{3}$. This effectively null result rules out appreciable anomalous entropy production during the high Mach number pericenter passage.

The finest cell sizes at the nozzle are $\Delta x=\Delta y=1.5625\times10^{-2}\,R_\odot$ and $\Delta z=7.8125\times10^{-3}\,R_\odot$. Direct half-maximum measurements from three meridional planes at fixed $\phi$ give $N_{\rm FWHM}=4.1$, $14.9$, and $15.4$ cells for the upstream, nozzle, and downstream profiles, respectively, i.e., physical widths of $0.032$, $0.117$, and $0.120\,R_\odot$.
These widths are consistent with defining $\phi_{\rm noz}$ by the minimum of $H_z$, because $H_z$ is dominated by the low-density envelope of the vertical profile, whereas the FWHM traces the dense core.

\begin{figure}
  \centering
  \includegraphics[width=\columnwidth]{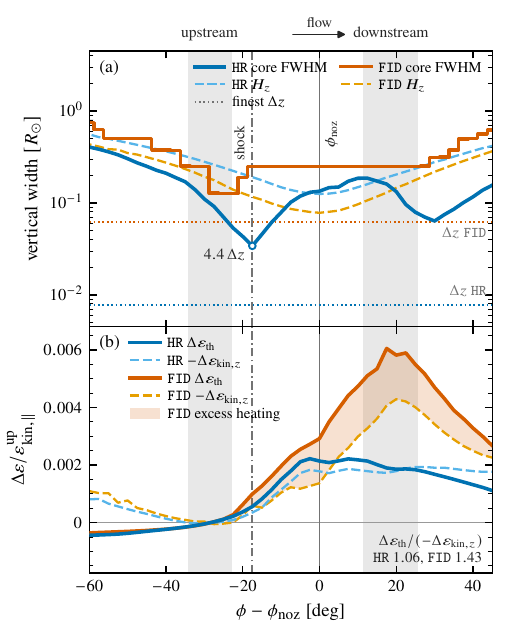}
  \caption{Collapse, shock, and vertical energy exchange along the stream at $t/P_{\rm mb}=1.70$ for \texttt{HR} (blue) and \texttt{FID} (orange), against the azimuth from $\phi_{\rm noz}$ of each run, with the flow from left to right. The dash-dotted line marks the shock at the minimum core FWHM of \texttt{HR}, and the grey bands are the windows of Table~\ref{tab:nozzle-resolution}. (a) Core FWHM (solid) and $H_z$ (dashed) of both runs. Dotted lines mark the finest $\Delta z$ of each run. The \texttt{FID} FWHM changes in steps of $\Delta z$ and stays at 2--4 cells near the shock, so its core is unresolved. (b) Thermal energy gained (solid) and vertical kinetic energy lost (dashed), relative to the upstream window and divided by $\varepsilon_{{\rm kin},\parallel}^{\rm up}$, with their window-averaged ratio. Shading marks \texttt{FID} heating in excess of the vertical energy loss.}
  \label{fig:nozzle-width-scan}
\end{figure}

Fig.~\ref{fig:nozzle-width-scan} follows both measures and the vertical energy budget along the flow. The vertical tidal field of the BH drives the stream to collapse toward the orbital plane, and the dense core reaches its minimum thickness first. The nozzle shock forms at this density peak, where the tidally driven vertical collapse of the core is halted. Behind the shock, the heated core re-expands, while the low-density envelope keeps converging and reaches its minimum thickness only at $\phi_{\rm noz}$. The core and the envelope therefore reach their minimum thickness at different azimuths. As a result, the upstream and nozzle planes of Fig.~\ref{fig:nozzle-energy-resolution} sample the core at maximum compression and during its re-expansion, respectively. A weaker second contraction about $30^\circ$ downstream of $\phi_{\rm noz}$ is likely a vertical oscillation of the shocked core, since its profile remains single peaked and the stream mass and radius vary smoothly. In all six snapshots of Table~\ref{tab:nozzle-resolution}, the specific entropy starts to rise about $20^\circ$ upstream of $\phi_{\rm noz}$, so the symmetric upstream window defined above contains the onset of the shock. For Fig.~\ref{fig:nozzle-width-scan} and Table~\ref{tab:nozzle-resolution}, we therefore average the upstream state over $0.40\le\phi_{\rm noz}-\phi\le0.60$~rad, which lies before the steep entropy rise, where the entropy has risen by less than a third of its total jump, and keep the downstream window. Ahead of the shock, the tidal field accelerates the gas vertically, and across the shock the thermal energy gained in \texttt{HR} matches the vertical kinetic energy lost (Fig.~\ref{fig:nozzle-width-scan}b). The nozzle heating is therefore the dissipation of the vertical collapse.

\begin{table*}
\centering
\caption{Nozzle energy budget of \texttt{FID} and \texttt{HR} at matched epochs, measured with identical windows. $\Delta s/s^{\rm up}$ is the relative increase of the specific entropy. $H_z$ and the core FWHM are given for the upstream window (up), $\phi_{\rm noz}$ (noz), and the downstream window (down). The finest $\Delta z$ is $7.8\times10^{-3}\,R_\odot$ in \texttt{HR} and $6.25\times10^{-2}\,R_\odot$ in \texttt{FID}. The upstream window lies ahead of the shock, $0.40$--$0.60$~rad upstream of $\phi_{\rm noz}$, so these values differ from the symmetric-window and single-plane values of Fig.~\ref{fig:nozzle-energy-resolution}.}
\label{tab:nozzle-resolution}
\footnotesize
\begin{tabular}{llcccccccccc}
\hline\hline
 & & & & & & \multicolumn{3}{c}{$H_z\,[R_\odot]$} & \multicolumn{3}{c}{FWHM $[R_\odot]$} \\
\cline{7-9}\cline{10-12}
$t/P_{\rm mb}$ & Run & $\Delta\varepsilon_{\rm th}/\varepsilon_{{\rm kin},\parallel}^{\rm up}$ & $-\Delta\varepsilon_{{\rm kin},z}/\varepsilon_{{\rm kin},\parallel}^{\rm up}$ & $\Delta\varepsilon_{\rm th}/(-\Delta\varepsilon_{{\rm kin},z})$ & $\Delta s/s^{\rm up}$ & up & noz & down & up & noz & down \\
\hline
1.70 & \texttt{HR}  & $1.93\times10^{-3}$ & $1.82\times10^{-3}$ & 1.06 & 0.14 & 0.272 & 0.125 & 0.201 & 0.101 & 0.134 & 0.139 \\
     & \texttt{FID} & $5.63\times10^{-3}$ & $3.93\times10^{-3}$ & 1.43 & 0.37 & 0.186 & 0.078 & 0.143 & 0.188 & 0.250 & 0.250 \\
1.94 & \texttt{HR}  & $4.26\times10^{-3}$ & $4.15\times10^{-3}$ & 1.03 & 0.30 & 0.420 & 0.135 & 0.258 & 0.101 & 0.055 & 0.055 \\
     & \texttt{FID} & $5.85\times10^{-3}$ & $5.29\times10^{-3}$ & 1.11 & 0.32 & 0.264 & 0.086 & 0.177 & 0.377 & 0.187 & 0.191 \\
2.17 & \texttt{HR}  & $6.56\times10^{-3}$ & $5.15\times10^{-3}$ & 1.27 & 0.39 & 0.563 & 0.271 & 0.373 & 0.117 & 0.139 & 0.071 \\
     & \texttt{FID} & $7.24\times10^{-3}$ & $5.36\times10^{-3}$ & 1.35 & 0.41 & 0.480 & 0.272 & 0.391 & 0.438 & 0.125 & 0.251 \\
\hline
\end{tabular}
\end{table*}

We repeat the same analysis for \texttt{FID}, whose finest cells are 4 times coarser in the orbital plane and 8 times coarser vertically (Table~\ref{tab:nozzle-resolution}). Run~\texttt{HR} is remapped from the shared state at $0.79\,P_{\rm mb}$, therefore the comparison isolates the resolution of the fallback stage. At its thinnest, the \texttt{HR} core spans only about 4 cells at the two earlier epochs, whereas near the shock the \texttt{FID} core stays at 2 to 4 cells at all epochs (Fig.~\ref{fig:nozzle-width-scan}a), a thickness set by the grid. Although this core is broader, \texttt{FID} has a smaller $H_z$ because its low-density envelope is weaker, holding $12\%$ of the upstream-window mass beyond $0.2\,R_\odot$ from the midplane, compared with $29\%$ in \texttt{HR}. At $t/P_{\rm mb}=1.70$, when only the thin tip of the most-bound debris has returned, the \texttt{FID} heating is $2.9$ times that of \texttt{HR} and exceeds the lost vertical kinetic energy by $43\%$ (Fig.~\ref{fig:nozzle-width-scan}b), whereas \texttt{HR} closes the budget to $6\%$. This excess is the numerical dissipation described by \citet{2026arXiv260520327N}. With the symmetric windows the ratio $\Delta\varepsilon_{\rm th}/(-\Delta\varepsilon_{{\rm kin},z})$ is $0.95$ in \texttt{HR} and $1.35$ in \texttt{FID}, so this contrast does not depend on the window placement. As the returning stream thickens, the heating in the two runs converges even though the \texttt{FID} core remains unresolved, and the difference drops to $10\%$ at $t/P_{\rm mb}=2.17$. There both runs heat more than the lost vertical kinetic energy, by $27\%$ in \texttt{HR} and $35\%$ in \texttt{FID}, so this excess is largely independent of resolution. At later times the outgoing stream crosses the analysis annulus and the diagnostic no longer isolates the nozzle. Strict convergence of the early compression remains beyond global three-dimensional calculations, and the one-dimensional simulations of \citet{2026OJAp....9E.106A} employ considerably higher resolutions.

\section{Radiative post-processing and synthetic observables}
\label{sec:radiative-postprocessing}

We construct synthetic observables from selected snapshots from the hydrodynamic simulations by adapting the photospheric post-processing method of \citet{2026ApJ...998..118Y}, originally applied to \texttt{Athena++} output. We summarize only the elements needed to interpret the observables here. 

Each image plane ray samples the leaf-level AMR solution along a specified line of sight by using the finest available MeshBlock at every location. The density and internal energy are mapped to the ray-sampling grid. The local temperature is then recovered from the same tabulated EOS used in the hydrodynamic evolution. The pipeline is adapted to Cartesian \texttt{AthenaK} AMR outputs and the H/He EOS of the simulation. The TDE simulation outputs do not contain the stellar passive scalar used by \citet{2026ApJ...998..118Y} to separate ejecta from the numerical atmosphere. We therefore exclude the BH excision region and numerical atmosphere before carrying out the ray integration.

Following the table-based option of \citet{2026ApJ...998..118Y}, we interpolate MESA opacities \citep{2011ApJS..192....3P,2013ApJS..208....4P,2015ApJS..220...15P,2018ApJS..234...34P,2019ApJS..243...10P} for $X=0.7$ and $Z=0.02$, which combine high-temperature OPAL opacities with low-temperature molecular and atomic opacities for the Grevesse--Sauval solar mixture \citep{1996ApJ...464..943I,1998SSRv...85..161G,2005ApJ...623..585F}. This solar-metallicity composition differs from the metal-free H/He EOS of the hydrodynamics and is an approximation confined to post-processing.

For a ray parameterized by path length $s$ toward an observer in direction $\hat{\bf n}$, the optical depth from $s$ to the observer $s_\mathrm{obs}$ is
\begin{equation}
  \tau(s,\hat{\bf n})
  =\int_s^{s_{\rm obs}}\kappa[\rho(s'),T(s')]\rho(s')\,ds' .
  \label{eq:rt-optical-depth}
\end{equation}
Here $\kappa$ is the Rosseland-mean total opacity of the MESA tables, including electron scattering, so the $\tau=1$ surface defined below is the scattering-inclusive photosphere and not the thermalization surface.
The photosphere on each ray is the first inward location at which $\tau=1$. Under the LTE photospheric approximation, its emergent specific intensity is
\begin{equation}
  I_\nu(\hat{\bf n})=B_\nu(T_{\rm ph}),
  \label{eq:rt-lte-intensity}
\end{equation}
where $T_{\rm ph}$ is the EOS recovered temperature at that crossing. Integrating over the projected image area gives the isotropic equivalent spectral and bolometric luminosities
\begin{align}
  L_{\nu,\rm iso}(\hat{\bf n})
  &=4\pi\int_{A_\perp} B_\nu(T_{\rm ph})\,dA_\perp,\\
  L_{\rm bol,iso}(\hat{\bf n})
  &=4\int_{A_\perp}\sigma_{\rm SB}T_{\rm ph}^4\,dA_\perp.
  \label{eq:rt-projected-luminosity}
\end{align}

We evaluate a face-on view with rays propagated from $+z$ toward $-z$ and an in-plane view from the $-y$ side toward $+y$, for which the nozzle lies on the near side of the flow. Because the photospheric temperature and projected area differ between the two orientations, both $L_{\nu,\rm iso}$ and $L_{\rm bol,iso}$ are view-dependent.

Relativistic Doppler boosting and aberration are neglected, because the maximum gas speed in the analyzed snapshot is only about $0.03c$.

The Rosseland mean does not locate the surface from which each frequency escapes, we also solve the transfer at 73 logarithmically spaced photon energies between $0.1$ and $1000$~eV. With the absorption and scattering coefficients $\alpha_{\rm a}(\nu)$ and $\alpha_{\rm s}=\sigma_{\rm T}n_e$, and $\epsilon=\alpha_{\rm a}/(\alpha_{\rm a}+\alpha_{\rm s})$, the emergent intensity is
\begin{equation}
  I_\nu=\int\frac{2\sqrt{\epsilon}}{1+\sqrt{\epsilon}}\,B_\nu\,e^{-\tau_*}\,d\tau_*,\,\,\,\,\,\,
  d\tau_*=\sqrt{\alpha_{\rm a}(\alpha_{\rm a}+\alpha_{\rm s})}\,ds,
  \label{eq:rt-thermalization}
\end{equation}
the modified-blackbody approximation for a homogeneous scattering atmosphere \citep{1979rpa..book.....R}, applied locally along each ray, which reduces to the formal LTE solution for $\epsilon\rightarrow1$. Here $\alpha_{\rm a}$ includes free--free absorption and the bound--free absorption of H, He\,\textsc{i}, and He\,\textsc{ii}, but no lines and no metals. The ionization fractions of H and He are interpolated in $\rho$ and $T$ from the tabulated EOS of the hydrodynamics (Sec.~\ref{sec:eos}), which at low density is the chemical-equilibrium EOS of \citet{2013ApJ...763....6T}, so the post-processing uses the same chemistry as the simulation. With $\alpha_{\rm a}$ replaced by the grey $\kappa\rho$, the scheme reproduces the grey formal solution to five digits, and doubling the number of photon energies changes the luminosity by less than $3\%$. We adopt this multifrequency transfer below and quote the grey photosphere for reference.

The snapshot-to-observable comparison previewed in Fig.~\ref{fig:tde-rt-multiview} shows that the photospheric emission retains the large-scale geometry of the dense fallback stream while preferentially highlighting the hotter material near the BH and nozzle. In the face-on view, the extended emitting surface follows the curved stream. The in-plane view compresses its vertically thin outer portion in projection and places the nozzle-side photosphere in the foreground. The luminosity therefore depends on the viewing angle. For the last \texttt{HR} snapshot ($t/P_{\rm mb}=2.93$), we obtain $L_{\rm bol,iso}=4.8\times10^{42}$ and $2.9\times10^{42}\,\mathrm{erg\,s^{-1}}$ for the face-on and in-plane views, with the spectrum peaking near $60$~eV. The grey $\tau=1$ photosphere gives $3.2\times10^{43}$ and $2.2\times10^{43}\,\mathrm{erg\,s^{-1}}$. It overestimates the luminosity by factors of 6.7 and 7.6 because the Rosseland mean is close to the electron-scattering opacity in the $\sim10^5$~K gas, whereas near the Planck peak the bound--free opacity is far larger and the emission emerges from cooler, outer layers. Doubling the image sampling from $512^2$ to $1024^2$ pixels changes the grey luminosities by $0.4\%$ and $6.1\%$.

These luminosities correspond to $\simeq33$ and $20\,L_{\rm Edd}$, with $L_{\rm Edd}\simeq1.5\times10^{41}\,\mathrm{erg\,s^{-1}}$ for $M_\bullet=10^3\,M_\odot$ and $\kappa_{\rm es}=0.34\,{\rm cm^2\,g^{-1}}$. At this epoch the fallback rate is $\dot M_{\rm fb}\simeq1.0\times10^{27}\,{\rm g\,s^{-1}}$, and the heating measured in the nozzle and inner self-intersection region, $\Delta\varepsilon_{\rm th}\simeq(6$--$8)\times10^{15}\,{\rm erg\,g^{-1}}$, supplies $(6$--$8)\times10^{42}\,\mathrm{erg\,s^{-1}}$, while the energy supply of the fallback, if all returning debris were circularized at $2r_{\rm t}$, is $GM_\bullet\dot M_{\rm fb}/(4r_{\rm t})\simeq5\times10^{43}\,\mathrm{erg\,s^{-1}}$. The multifrequency luminosity is therefore comparable to, if somewhat below, the instantaneous dissipation rate, while the grey photosphere exceeds it by a factor of 3--5. Both remain illustrative post-processing estimates. The hydrodynamic calculation is adiabatic and omits radiative cooling, which is not negligible when the emitted power is comparable to the dissipation rate, no radiative energy budget is enforced, photon trapping and advection are neglected, and the result depends on the density threshold that separates the debris from the numerical atmosphere, most strongly in the in-plane view. A prediction of the emergent luminosity at these super-Eddington rates requires radiation hydrodynamics.

Applying the same procedure to a time-ordered sequence of snapshots yields viewing-angle-dependent light curves without modifying the hydrodynamic calculation. We do not present light curves here, because the simulation does not reach the peak of the flare, and defer them to future work.

\section{Summary and Limitations}
\label{sec:summary-limitations}

We have built an end-to-end TDE framework in AthenaK that follows a star from hydrostatic equilibrium through disruption and fallback in a unified simulation, with the gas self-gravity, the BH potential, the frame state, and a tabulated H/He EOS carried consistently throughout. The multigrid solver runs 4.4 times faster than the public AthenaK solver on the same test, and the debris energy distribution changes by at most 4\% when its refresh interval is reduced twelvefold. A moving frame with repeated domain remaps handles the growth of the debris, and the remaps carry the thermal rather than the total energy, so that kinetic energy unresolved on the coarser grid is not converted into heat in a stream whose thermal energy is only $10^{-3}$ of its kinetic energy. The dual-energy formulation keeps the cold stream thermodynamically correct at Mach numbers over $10^{4}$, where the conservative subtraction fails, and LAT accelerates the production run by a factor of 3.1. Each module is validated against controlled tests, and the fiducial $10^{3}\,M_\odot$ calculation tests them in combination.

The pericenter nozzle is the most demanding part of the flow. The vertical tidal field of the BH drives the returning stream to collapse toward the orbital plane, and the nozzle shock forms where this collapse of the dense core is halted, about $17^\circ$ upstream of the minimum of the mass-weighted thickness. Even at $r_{\rm t}/\Delta z=1280$, the shock is resolved by only a few cells. A run with 4 to 8 times coarser cells overheats the earliest, thinnest stream by a factor of 3, which is the numerical dissipation identified by \citet{2026arXiv260520327N}, whereas at the higher resolution the heating matches the lost vertical kinetic energy to 6\%, and the two resolutions agree to 10\% once the stream has thickened. The debris energy distribution has the expected width and an even split into bound and unbound mass but is not flat, and its bound side reproduces the late-time $t^{-5/3}$ fallback. For the synthetic observables, a grey Rosseland photosphere overestimates the luminosity by about a factor of seven relative to a frequency-dependent continuum calculation, because near the Planck peak the bound--free opacity moves the emitting surface to cooler, outer layers.

The presented methodology relies on several explicit physical and computational boundaries. The primary calculation explores a single Newtonian encounter with $\beta=1$ \citep[see][for relativistic effects]{2020ApJ...904..101R,2026ApJ..1003..221C}. The energy-inferred fallback curves are not direct sink accretion rates. The remapping scheme projects conserved variables but is not an exact geometric overlap method \citep{1987SJSSC...8..305D}. It carries the thermal energy of the gas and rebuilds the total energy from it, so the kinetic energy that a coarser grid cannot represent is dropped rather than turned into heat, and the total energy is not conserved across a remap by construction. Its accuracy is established primarily by the reported checkpoint audits. Furthermore, the radiative calculation is an LTE post-processing model, and its luminosities of $\simeq20$--$33\,L_{\rm Edd}$ are illustrative, since no radiative energy budget is enforced. It does not evolve radiation energy or momentum with the gas, as radiation-hydrodynamic calculations of disk formation do \citep{2026ApJ..1009...64H}. It omits non-LTE source functions, scattering-dominated spectral formation, and radiative feedback on the dynamics. Additionally, the solar metallicity opacity table used for post-processing is compositionally inconsistent with the metal-free hydrodynamic EOS. The multifrequency transfer also omits the bound-free and line opacity of metals, which would add absorption in the extreme ultraviolet near the spectral peak.

Despite these recognized limitations, the framework provides a tested route from stellar initialization to late AMR snapshots and approximate synthetic observables. It establishes a robust numerical foundation while keeping validation claims strictly separate from the physical interpretation of circularization detailed in the companion paper \citep{Jiang2026Science}. Future methodological upgrades will incorporate exact conservative geometric remapping, broader multi-resolution production suites, and compositionally consistent opacities. Future work will extend LAT to support the additional physics modules needed for TDE modeling. These include radiation hydrodynamics with an M1 closure in both Newtonian and general relativistic frameworks, two-temperature thermodynamics \citep{2025ApJ...995..112J,2025ApJ...990...81J,2023MNRAS.522.2307J}, and magnetic fields on dynamic spacetime backgrounds \citep{2026ApJ..1006..128J}.

Code availability: The code used in this work is publicly available at \url{https://github.com/HongxuanJiang/athenak-transients}. The release contains the TDE problem generator, example input files for the calculations presented here, and the radiative post-processing package.

\begin{acknowledgments}
The authors thank Zachary L. Andalman for discussions on EOS. YM is supported by the National Key Research
and Development Program of China (grant no. 2023YFE0101200), the National Natural Science Foundation of China (grant nos.12273022, 1251154005, W2641030), and the Shanghai Municipality Orientation Program of Basic Research for International Scientists (grant no.22JC1410600).
\end{acknowledgments}

\bibliography{sample701}{}
\bibliographystyle{aasjournalv7}

\end{document}